\documentclass[11pt]{article}
\usepackage{graphicx}
\usepackage[numbers]{natbib}

\makeatletter

\usepackage{amsmath,multirow}
\usepackage{enumerate,array}
\usepackage{amsfonts}
\usepackage{mathtools}
\usepackage{amssymb}
\usepackage{epsfig}
\usepackage{graphics}
\usepackage{multicol, multirow}
\usepackage[pdfstartview=XYZ, bookmarks=true, colorlinks=true, linkcolor=blue, urlcolor=blue, citecolor=blue, pdftex, bookmarks=true, linktocpage=true, hyperindex=true]{hyperref}
\usepackage{enumitem}
\usepackage{color}
\usepackage{amsthm}
\usepackage{float}
\usepackage{lscape}
\usepackage{verbatim}
\usepackage{amsfonts}
\usepackage{lineno}
\usepackage[numbers]{natbib}
\makeatletter

\newcommand{\beq}{\begin{equation}}
\newcommand{\eeq}{\end{equation}}
\newcommand{\ux}{\: \underline{x}}
\newcommand{\uw}{\: \underline{w}}
\newcommand{\uwh}{\: \hat{\underline{w}}}
\newcommand{\ur}{\: \underline{r}}
\newcommand{\siK}{\sum_{k=1}^K}

\makeatletter
\newcommand{\pushright}[1]{\ifmeasuring@#1\else\omit\hfill$\displaystyle#1$\fi\ignorespaces}
\newcommand{\pushleft}[1]{\ifmeasuring@#1\else\omit$\displaystyle#1$\hfill\fi\ignorespaces}
\makeatother

\renewcommand{\baselinestretch}{1.2}

\begin{document}
	
\title{Model averaging with mixed criteria for estimating \\ high quantiles of extreme values: Application to heavy rainfall}

   \author{Yonggwan Shin$^{1}$,~Yire Shin$^{2, *}$,~ Jeong-Soo Park$^{2}$  \\	\\
   \small \it 1: 
   Honam Research Division, Electronics and Telecommunications Research Institute,\\ Gwangju, 61012, Korea
   \\ 
   \small \it 2: Department of Statistics, Chonnam National University, Gwangju 61186, Korea \\
   \small \it *: Corresponding author, e-mail: shinyire87@gmail.com 
}
\footnotetext[1]{This is a preprint version of the published paper: Shin, Y., Shin, Y. \& Park, JS. Model averaging with mixed criteria for estimating high quantiles of extreme values: application to heavy rainfall. Stoch Environ Res Risk Assess 40, 47 (2026). https://doi.org/10.1007/s00477-025-03167-x
}
   \maketitle  
	
	\begin{abstract}
 Accurately estimating high quantiles beyond the largest observed value is crucial for risk assessment and devising effective adaptation strategies to prevent a greater disaster. The generalized extreme value distribution is widely used for this purpose, with L-moment estimation (LME) and maximum likelihood estimation (MLE) being the primary methods. However, estimating high quantiles with a small sample size becomes challenging when the upper endpoint is unbounded, or equivalently, when there are larger uncertainties involved in extrapolation. This study introduces an improved approach using a model averaging (MA) technique. The proposed method combines MLE and LME to construct candidate submodels and assign weights effectively. The properties of the proposed approach are evaluated through Monte Carlo simulations and an application to maximum daily rainfall data in Korea. In addition, theoretical properties of the MA estimator are examined, including the asymptotic variance with random weights. A surrogate model of MA estimation is also developed and applied for further analysis. {Finally, a Bayesian model averaging approach is considered to reduce the estimation bias occurring in the MA methods.}


\end{abstract}
	
\vspace{5mm} \noindent {\bf Keywords}: {Bootstrap, 
	 generalized L-moment distance, profile likelihood, restricted maximum likelihood estimation,
	 return level, surrogate model.}

	\section{Introduction}
	
The generalized extreme value (GEV) distribution (GEVD) is extensively used to model extremes in natural phenomena and human activities, such as hydrological events and incidents in insurance and financial markets (Coles 2001; Katz et al.~2002; Reiss \& Thomas 2007). 
 The primary reason for applying the GEVD is that it serves as a large-sample approximation to the distribution of sample maxima, regardless of the population from which the data are drawn. 
The cumulative distribution function (CDF) of the GEVD is given by (Hosking \& Wallis 1997):
\begin{linenomath*}\begin{equation} \label{cdf-gevd}
	F(x) = \text{exp} \left\{ - \left(1 -\xi {\frac{x-\mu}{\sigma}}\right) ^{1/ \xi} \right\}.
	\end{equation}\end{linenomath*}
 when $1-\xi (x- \mu ) / \sigma > 0$ and $\sigma>0$, where $\mu,\; \sigma $, and $\xi$ represent the location, scale, and shape parameters, respectively. The case where $\xi=0$ in (\ref{cdf-gevd}) corresponds to the Gumbel distribution. 
 Notably, the sign of $\xi$ in (\ref{cdf-gevd}) differs from that used in Coles (2001) to align with the notation in Hosking \& Wallis (1997). The tail behavior of the GEV distribution varies depending on the sign of $\xi$: it exhibits a heavy tail when $\xi<0$, an exponential tail as $\xi \rightarrow 0$, and light or bounded when $\xi>0$. 

To estimate the three parameters of the GEVD, maximum likelihood estimation (MLE) and L-moment estimation (LME) methods are commonly used. The LME generally performs well with small samples and offers advantages over the MLE, including computational simplicity, greater robustness, and lower estimation variance (Hosking \& Wallis 1997; Morrison \& Smith 2002; Karvanen 2006; Delicado \& Goria 2008). Conversely, the MLE is preferable for large samples due to its desirable asymptotic properties. 

 Accurately estimating high upper quantiles is more critical than estimating the parameters of extreme value distributions, particularly in risk assessment and the design of hydraulic structure  (Naghettini 2017; Allouche et al.~2023; Zheng et al.~2025). For example, the 1\% chance flood is computed at the 0.99 quantile of an extreme value distribution which is actually used by the US Federal Emergency Management Agency when developing its regulatory policy for floodplains (Grego \& Yates 2025). When the upper endpoint is unbounded (i.e., $\xi <0$), estimating high quantiles of the GEVD with a small sample size becomes challenging because of larger uncertainties associated with model extrapolation. For instance, when $\xi= -0.35$ and the true $0.99$ quantile is 443 (with $\mu=100$ and $\sigma=30$), the root mean squared errors (RMSEs) of the MLE and LME of the $0.99$ quantile with $n=60$ are approximately 148 and 120, respectively. These RMSEs represent 33\% and 27\% of the true quantile, which is considered a substantial estimation uncertainty.

In the context of the GEVD for block maxima data, several approaches have been proposed to address this issue, including a mixed estimation of the MLE and LME (Morrison \& Smith 2002; Ailliot et al.~2011) and the application of a penalty function to the GEVD parameters (Coles \& Dixon 1999; B{\"u}cher et al.~2021). The former method estimates the parameters effectively and robustly while reducing RMSE, whereas the latter can outperform MLE in certain situations depending on the choice of penalty function.
 {Lekina et al.~(2014) and Fabiola et al.~(2021) introduced semi-parametric and nonparametric weighted estimations of high quantiles, respectively, offering more flexibility and suitability than parametric methods. Daouia et al.~(2024) presented general weighted pooled estimators of extreme quantiles calculated through a nonstandard geometric averaging scheme. These weighted estimation methods generally perform well for large sample sizes.}
 Karvanen (2006) considered the estimation of quantile mixtures via L-moments and trimmed L-moments. Grego et al.~(2015) used finite mixture models to accommodate annual peak streamflow. Evin et al.~(2011) and Totaro et al.~(2024) explored the use of two-component mixtures of Gumbel distributions. Our study builds upon this framework by introducing a modified approach that employs mixtures of narrow extreme value submodels.

In this study, we propose a new estimation method aimed at improving the accuracy of high quantile estimation in the GEVD using model averaging (MA). The MA combines the strengths of multiple candidate models by assigning greater weights to superior models (Liu et al.~2016; Fletcher 2018). Unlike traditional model selection, which discards all but the best model, MA explicitly accounts for model uncertainty by integrating information from competing statistical models. 
MA reduces the risk of model misspecification and makes predictions less sensitive to model assumptions. By incorporating model uncertainty, MA produces more accurate estimates and more reliable confidence intervals than model selection from frequentist and Bayesian perspectives (Burkland et al.~1999; Hoeting et al.~1999; Claeskens \& Hjort 2008). 
{Le and Clarke (2022) showed that MA is asymptotically better than model selection in terms of  predictive performance.}

However, the MA method has some limitations, such as computational cost. The slowest candidate model largely determines the overall computational speed of MA.
Additionally, MA does not consistently outperform model selection. Model selection is preferred when the error variance is small and identifying the best model is relatively straightforward. Conversely, MA performs well when the error variance is large (Liu et al.~2016, 2023). Given the high error variance in estimating high quantiles of the GEVD under a negative shape parameter, MA is expected to outperform some existing methods. 

Several Bayesian MA approaches for extreme values have been developed based on the likelihood
function and posterior probability (Raftery et al.~2005; Sabourin et al.~2013; Vettori et al.~2020). These approaches typically estimate the parameters of candidate models using MLE or Bayes estimators and assign weights based on the marginal likelihood. In frequentist MA, candidate model parameters are estimated using the least squares method or MLE, and weights are determined based on information criteria, cross-validation, or an unbiased risk estimator (Claeskens \& Hjort 2008; Wang et al.~2009). 

This study applies different criteria for estimation and weighting in MA, with a primary focus on  L-moments. This mixed application of criteria is a key aspect of the proposed method, which was motivated by Morrison \& Smith (2002) and Ailliot et al.~(2011). 
We hypothesize that the mixed-criteria approach—where candidate model parameters are estimated using MLE and weights are assigned based on L-moments, or vice versa—can outperform the conventional single-criterion method, which estimates parameters via MLE and assigns weights based on the likelihood function. We anticipate that this mixed-criteria approach will improve the estimation of high quantiles of extreme values. To our knowledge, L-moment-based MA methods have been rarely studied, despite the popularity of non-MA approaches based on L-moments in climatology and hydrology (Hosking \& Wallis 1997; Naghettini 2007; Kjeldsen et al.~2017; Lee et al.~2020; Strong et al.~2025, among others).

 We consider two methods: (1) estimating candidate model parameters using MLE and calculating weights using the generalized L-moment distance, and (2) estimating candidate model parameters using LME and calculating weights using the smooth Akaike information criterion. For comparison, we also evaluate other methods, including mixed estimation (Morrison \& Smith 2002; Ailliot et al.~2011) and a penalized likelihood method (Coles \& Dixon 1999). 

 The rest of this paper is structured as follows: 
 Section 2 details the proposed MA methods. Theoretical considerations on the MA estimator and uncertainty calculation under random weights are presented in Section 3. Section 4 provides a simulation study. Section 5 presents an application using maximum rainfall data from Hae-nam, Korea. Further development including the selection of the number of submodels and bias correction using Bayesian model averaging are presented in Section 6. Discussion and conclusion are provided in Sections 7 and 8. The Supplementary Material includes additional details such as formulas, tables, and figures. The R code for our implementation is available on GitHub: https://github.com/yire-shin/MA-gev.git.

\section{Model averaging methods}

The MA methodology consists of three steps: (1) constructing candidate models, (2) estimating model parameters, and (3) calculating weights for each candidate model (Liu et al.~2023; Salaki et al.~2024; Hao et al.~2024). The primary objective of MA in this study is to improve the estimation of high quantiles of extreme values, rather than to focus on parameter estimation of the GEVD. The $1/p$ return level, denoted as $r_{1/p}$, represents the $1-p$ quantile of the distribution (Coles 2001). The ``T-year return level” is commonly used for annual extreme values, where $T=1/p$. The MLE and LME of the three GEVD parameters are defined as follows:
\beq 
\hat{\theta}_\text{M}= (\hat{\mu}_\text{M}, \hat{\sigma}_\text{M}, \hat{\xi}_\text{M}),
~~~~
\hat{\theta}_\text{L}= (\hat{\mu}_\text{L}, \hat{\sigma}_\text{L}, \hat{\xi}_\text{L}).
\eeq

\subsection{Construction of candidate models}

A concrete application of MA often requires an initial selection of several candidate models. These candidate models may differ entirely or only in certain components. 
Among the three parameters in the GEVD, estimating the shape parameter $\xi$ is more challenging than estimating the other two parameters. Additionally, $\xi$ is dimensionless, and the behavior of the GEVD is more sensitive to $\xi$ than to the other two parameters. Therefore, the proposed method starts with the shape parameter $\xi$. We consider two approaches for initializing our MA algorithm: the MLE of $\xi$ and the LME of $\xi$.

\begin{figure}[h!bt]
	\centering
	\begin{tabular}{c}	\includegraphics[width=13cm, height=14cm]{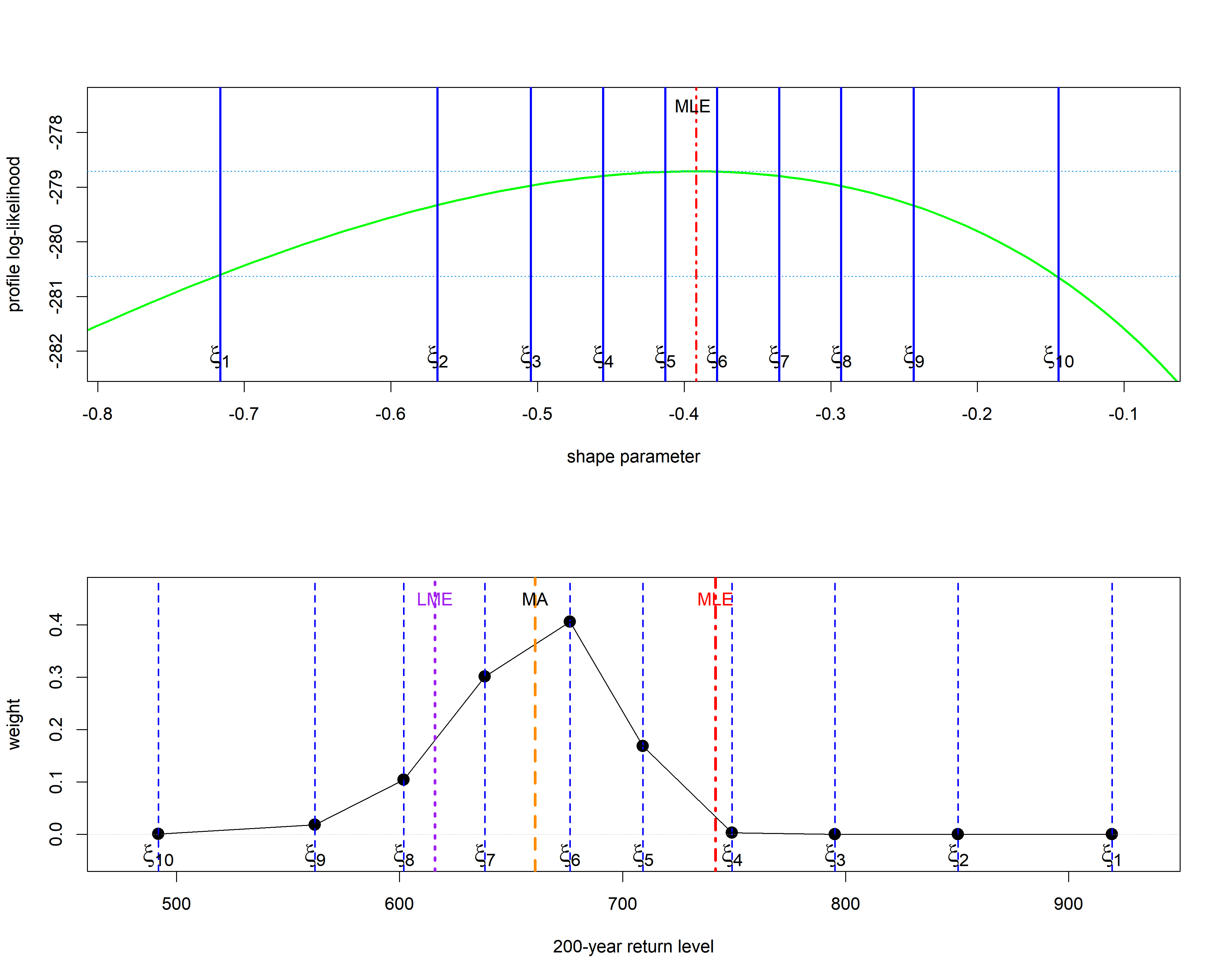}
	\end{tabular}
	\caption{Illustration of the selection of $\xi_k$ values using the profile likelihood of $\xi$ and the corresponding weights, based on annual maximum daily rainfall data (unit: $mm$) from Hae-nam, Korea. (Top panel): The green curve represents the profile log-likelihood function, while the ten  vertical lines indicate the selected $\xi_k$ values. (Bottom panel): The 200-year return level estimates from ten submodels (vertical dotted lines) are shown along with their corresponding weights (black dots) for the proposed MA method (`like1'). The return level estimates obtained from the MLE, LME, and MA methods are also displayed for comparison.
	} \label{Hae-nam_kpar}
\end{figure}

In the first approach, based on $\hat{\xi}_M$, we construct a $100 \times (1-\alpha)$ confidence interval for $\xi$ using the profile likelihood (Coles 2001). For instance, setting $\alpha=0.05$ provides a 95\% confidence interval $(C_L,\; C_U)$. We select $K$ values of $\xi$ proportionally to the profile likelihood of $\xi$ within this interval from $C_L$ to $C_U$, since values near the lower and upper bounds are less probable than in the interior of the interval. The upper panel of Figure \ref{Hae-nam_kpar} illustrates how the proposed method selects ten values of $\xi_k$ using a 95\% confidence interval of $\xi$ based on the profile likelihood function.
In the second approach, based on $\hat{\xi}_L$, we use a nonparametric bootstrap to construct a $100 \times (1-\alpha)$ confidence interval for $\xi$. Subsequently, we select $K$ values of $\xi$ with equal probability from the bootstrap distribution of $\hat{\xi}_L$ within this confidence interval. This study considers two values of $\alpha$: $0.05$ and $0.10$. 

 For each selected $\xi_k$, we estimate $\mu_k$ and $\sigma_k$  using either the MLE or LME, keeping $\xi_k$ fixed. These estimators are denoted as $\hat{\mu}_{k \text{M}},\; \hat{\sigma}_{k \text{M}}$, and  $\hat{\mu}_{k \text{L}},\; \hat{\sigma}_{k \text{L}}$, respectively. Estimating the two other parameters with $\xi_k$ fixed is computationally simpler and yields lower standard errors (SE) than estimating all three GEVD parameters. When $\xi$ is known, the LMEs of $\mu_k$ and $\sigma_k$  are given by:
 \beq \label{lme_xifix}
 \begin{aligned}
 &\hat{\sigma_k} = \frac{l_2 \ \xi_k }{(1-2^{\xi_k} )\ \Gamma(1+\xi_k)}, \\
 &\hat{\mu_k} = l_1 - \hat{\sigma_k}\ \{ 1- \Gamma(1+ \xi_k)\}/ \xi_k,
 \end{aligned}
 \eeq
 where $l_1$ and $l_2$ are the first and second sample L-moments computed from the data.
As a result, we obtained $K$ GEV candidate models, where each is a two-parameter GEVD. From each candidate model, we compute the T-year return level ($\hat{r}_{T,k}$) using the estimated parameters $(\hat{\mu}_{k \text{M}},\; \hat{\sigma}_{k \text{M}},\; \xi_k)$ or $(\hat{\mu}_{k \text{L}},\; \hat{\sigma}_{k \text{L}},\; \xi_k)$. 
Finally, we compute the MA T-year return levels by averaging these T-year return levels with weights $w_k$, which are nonnegative and sum to one.

\subsection{Weighting scheme based on L-moments distance} \label{sec:w_L-moment}
 
When the parameters in each candidate model are estimated using the MLE, reusing the likelihood value to compute weights—such as through the smooth Akaike information criterion (AIC) (Claeskens \& Hjort 2008; Fletcher 2018)—may not necessarily improve estimation, in certain settings. As demonstrated later in this study, conventional model averaging that employs MLE for parameter estimation and smooth AIC for weight assignment does not yield satisfactory performance, at least in our simulation study. Therefore, we explored alternative criteria to construct weights that are more informative for high-quantile estimation.

For a three-parameter distribution, the L-moment distance for each candidate model $M_k$ is defined as follows, subject to the constraint that $\xi > -1$ in the GEVD:
\beq \label{L-dist1}
LD_k = | l_1 -\lambda_{1, k} | + |l_2 - \lambda_{2,k} | + |l_3 -\lambda_{3,k} |,
\eeq
where $l_j$ represents the $j$th sample L-moment, and $\lambda_j$ denotes the $j$th population L-moment, for $j=1, 2, 3$. The sample L-moments are computed directly from the data, whereas $\lambda_{j, k}$ values are derived from the GEVD using the estimated parameters for each candidate model $M_k$. 
 L-moments offer both theoretical and practical advantages over conventional moments. Specifically, they exhibit lower sensitivity to outliers and provide more reliable identification of the parent distribution generating the data (Hosking \& Wallis 1997; Lomba \& Alves 2020). 

Since the LME is obtained by solving equations that ensure $LD_k =0$, we aim to assign higher weights to submodels with smaller L-moment distances. This weighting scheme is preferable because L-moments characterize the distribution differently from the likelihood function.

 One challenge in using the L-moment distance is that the three L-moments operate on different scales. Typically, the first term in (\ref{L-dist1}) dominates the distance metric, while the third term —corresponding to the crucial shape parameter—is nearly negligible. To address this issue, we introduce a generalized L-moment distance (GLD) for each model $M_k$:
\beq \label{GL-dist}
\text{GLD}_k = d_k^T \ V_k^{-1} d_k,
\eeq
where $d_k^T = ( l_1 -\lambda_{1, k} ,\; l_2 - \lambda_{2,k},\; l_3 -\lambda_{3,k} )$, and $V_k$ is a $3 \times 3$ covariance matrix of $d_k$. A similar GLD approach has been explored in previous studies, including Kjeldsen \& Prosdocimi (2015) for regional frequency analysis as well as Alvarez et al.~(2022), Shin \& Park (2024), and Shin et al.~(2025a). 

To compute $V_k$ for each model, one can use a parametric bootstrap based on the GEVD with parameter estimates $(\hat{\mu}_{k \text{M}},\; \hat{\sigma}_{k \text{M}},\; \xi_k)$ since an explicit form of $V_k$ is unavailable.
However, this approach is computationally demanding. To mitigate this issue, we approximate $V_k$ using the covariance matrix of the sample L-moments (Elamir \& Seheult 2004), treating $\lambda_{j, k}$ as constants. For this computation, we employed the `lmomco' package (Asquith 2014) in R. However, in some cases, this method resulted in a non-positive definite covariance matrix. In such instances, we computed the covariance matrix using a nonparametric bootstrap. Once computed, $V$ was fixed and applied uniformly across all $K$ candidate models, as it no longer depended on $k$. 

To assign higher weights to submodels with smaller $\text{GLD}_k$ values, we adopted the multivariate normal density:
\beq \label{probMN}
p(\ux | M_k) ={ \frac{1}{(2 \pi )^{3/2} |V|^{1/2}} } \; \text{exp} (- d_k^T \ V^{-1} d_k /2 ) .
\eeq
The weight for model $M_k$ is calculated as follows:
\beq \label{weight}
\hat w_k ={ \frac{p(\ux |M_k)}{\sum_{j=1}^K p(\ux | M_j) } },
\eeq
which ensures that $\sum_{k=1}^K \hat w_k = 1$ and that all $\hat w_k$ values remain within the range $[0,1]$.
This weight can be interpreted as the probability that model $M_k$ is the best among the set of candidate models. Given that sample L-moments generally converge to a multivariate normal distribution as $n \rightarrow \infty$ (Hosking 1990), the function in (\ref{probMN}) serves as an approximation of the likelihood function of population L-moments given three sample L-moments. 

For a more robust L-moment distance, one can replace $l_1$ and $l_2$ with the sample median and interquartile range of the data. While $l_2$ is proportional to Gini's mean difference and thus already serves as a robust scale estimate, $l_1$ (the sample mean) is not (Ailliot et al.~2011). To address this, we substitute the first element of $d_k$ with the median-based distance ($\text{sample median} - q_{0.5}$) as an alternative to (\ref{GL-dist}). In the following sections, we refer to this alternative weighting scheme as the `med’ weight.

The lower panel of Figure \ref{Hae-nam_kpar} illustrates how the proposed method computes the return level estimate using a weighted average of return values from ten submodels. The 200-year return level obtained via the MA method lies between the estimates derived from the MLE and LME.

\subsection{Estimation of high quantiles and building a surrogate model}

Our MA method estimates the high quantile or T-year return level using the following formulation:
\beq \label{rlMA}
\hat{r}_\text{MA} = \sum_{k=1}^K \hat w_k \; \hat{r}_{k},
\eeq
which is a weighted average of the return level predictions from individual submodels.
We aim to achieve this with lower estimation uncertainty than the MLE, LME, and other established methods.

The limitation of this MA approach is that it may not be useful for system description and understanding (Fletcher 2018). We consider constructing a weighted average (or ensemble) model from the $K$ GEV candidate models to address this limitation. However, the ensemble model may not necessarily follow a GEVD, even though each model $M_k$ does. Thus, simply taking a weighted sum of $K$ GEV candidate models is not an ideal solution. Instead, we propose building a surrogate model that still follows a GEVD and that best approximates the return level estimates generated by the MA method.

 To achieve this, we estimate the three GEVD parameters by minimizing the following function with respect to $\mu$, $\sigma$, and $\xi$:
\beq
RSS(\mu, \sigma,\xi) = \sum_{i=1}^I (r_{q_i} - \hat r_{q_i})^2 ,
\eeq
where $\{r_{q_i}\}_{i=1,\dots, I}$ are the return levels corresponding to probability $q_i$, computed from a $GEVD(\mu, \sigma,\xi)$, and $\hat r_{q_i}$ are the corresponding return level estimates obtained using the MA method. To ensure a broad coverage of important quantiles, we selected $I=16$ probabilities ranging from $0.5$ to $0.999$.
 Although the surrogate model is only an approximation and may not be theoretically exact, it is computationally convenient and useful for further data analysis. This convenience includes additional predictions, model comparison, and model diagnostics such as quantile-per-quantile and return level plots. 
 Of course, other models could also be used as surrogates instead of using the GEVD.

\section{Theoretical consideration}
\subsection{Conditional expectation}

We provide a theoretical justification for why the proposed MA method is expected to perform well.
If we assume that $r(\hat \mu, \hat \sigma, \hat \xi)$ is an unbiased estimator of the return level $r(\mu,\sigma,\xi)$, then we have:
\beq \label{condexp}
r(\mu,\sigma,\xi)\ =\ E \: \{ r(\hat \mu, \hat \sigma, \hat \xi)  \}\ =\ E_{\xi} \: \left[ E \: \{r(\hat \mu, \hat \sigma, \hat \xi) \: | \: \hat \xi = \xi \} \right].
\eeq
If $E \{r(\hat \mu, \hat \sigma, \hat \xi) | \hat \xi = \xi \}$ serves as an estimator of $r(\mu,\sigma,\xi)$, then it is unbiased and has a smaller variance than $r(\hat \mu, \hat \sigma, \hat \xi)$, as follows from the conditional variance identity (Casella \& Berger 2001),
\beq \label{condvar}
\begin{aligned}
Var \{ r(\hat\mu, \hat\sigma, \hat\xi) \} & \ =\ Var \left[  E \{r(\hat\mu, \hat\sigma, \hat\xi)  \:| \: \hat\xi = \xi \}  \right] \: + \: E \left[ Var \{ r(\hat\mu, \hat\sigma, \hat\xi)  \:|  \: \hat\xi = \xi  \} \right] \\
&\ \ge \ Var \left[  E \{ r(\hat\mu, \hat\sigma, \hat\xi) \: | \: \hat\xi = \xi \}  \right].\\
\end{aligned}
\eeq
If $\hat\xi$ is a sufficient statistic for $r(\mu,\sigma,\xi)$, then $E \{r(\hat \mu, \hat \sigma, \hat \xi)\, |\, \hat \xi = \xi \}$ is a uniformly better unbiased estimator, as stated by the Rao-Blackwell theorem (Casella \& Berger 2001). However, $E \{r(\hat \mu, \hat \sigma, \hat \xi)\, |\, \hat \xi = \xi \}$ is not an estimator, as it depends on the unknown parameter $\xi$. 
To address this, we approximate $E \{r(\hat \mu, \hat \sigma, \hat \xi)\, |\, \hat \xi = \xi \}$ using a weighted average of available estimators computed for $K$ values of $\xi$s near $\hat \xi$, leveraging relatively simple estimation methods. 

Since estimating $\xi$ is more challenging than estimating the other parameters of the GEVD and because $\hat \xi$ has the most substantial influence on the estimation of $r(\mu,\sigma,\xi)$—particularly when $\xi < -0.2$—conditioning on $\xi$ as a fixed value near $\hat \xi$ can substantially simplify the estimation process. The generally lower variability of a narrow submodel compared to a broader model is well demonstrated in Claeskens \& Hjort (2008) under the local misspecification framework. Additionally, fitting submodels is computationally more efficient than fitting a full model.

Assuming that $(\hat \mu, \hat \sigma, \hat \xi)$ can always be estimated  from the data, a heuristic derivation of $\hat{r}_\text{MA}$ proceeds as follows: 
\beq \label{heuri-1}
\begin{aligned}
 & E \left[\: r\{\hat \mu, \hat \sigma, \hat \xi \} \: |\: \hat \xi = \xi\, \right]  
 \ =\: E \left[\: r\{ \hat \mu_2(\xi), \hat \sigma_2(\xi), \xi \} \: | \: \hat \xi = \xi\, \right]    \\ %
  & ~~~~~~~~~~~~~~~~~~~~~~~~~~~~~~~~~~~~~~~~~~~~~~~~~~~~~~ \begin{pmatrix}  \hat \mu_2(\xi), \hat \sigma_2(\xi) \text{ are two-dimentional } \\  \text{estimators for a given constant}\ \xi \end{pmatrix}  \\
&\ =\: \int  r\{\hat \mu_2(\xi), \hat \sigma_2(\xi), \xi \} \times p\left[\, r\{ \hat \mu_2(\xi), \hat \sigma_2(\xi), \xi \} \: | \: \hat \xi = \xi \, \right] \; dr \\
&~~~~~~~~~~~~~~~~~~~~~~~~~~~~~~~~~~~~~~~~~~~~~~~~~~~~~~ \begin{pmatrix} \hat \mu_2(\xi),\: \hat \sigma_2(\xi) \text{ are no longer random} \\
	\text{ for a given constant $\xi$ and the data} \end{pmatrix} \\
&\ =\: r\{ \hat \mu_2(\xi), \hat \sigma_2(\xi), \xi \} \\
&\ \xleftarrow{\text{  \large estimate  }} \: \sum_{k=1}^{K}\: r \{ \hat \mu_2(\xi_k), \hat \sigma_2(\xi_k), \xi_k \} \times \hat w (\xi_k), 
  \end{aligned}
  \eeq
  where $\xi_k$ values are selected based on the sampling distribution of $\xi$. Thus, our MA method can be interpreted as approximately integrating out the $\xi$ parameter. 

 \subsection{Asymptotic distribution of MA return level}
 
 When submodel $M_k$ is fitted using the MLE, the MLE of $(\mu_k,\ \sigma_k)$ for a given $\xi_k$ follows an asymptotically normal distribution with mean $(\mu_k,\; \sigma_k)$ and covariance matrix $\Sigma_k$, which is the inverse of the Fisher information matrix, provided that $-1 < \xi_k < 1/2$ (B{\"u}cher et al.~2021, for example). Consequently, $\hat r_k$ asymptotically follows a normal distribution with mean $r_k$ and variance:
 \beq \label{delta1}
 Var(\hat r_k)\  \approx\ \nabla r_k^T \;\Sigma_k \;\nabla r_k ,
 \eeq
 by the delta method, where $\nabla r_k$ is a gradient vector of $r_k$ with respect to $(\mu_k,\; \sigma_k)$. 
 For the GEV distribution, the gradient vector of $r_k$ with respect to $(\mu_k,\; \sigma_k)$ is given by
 \beq
 \nabla r_k\, =\, [1,\; \xi_k^{-1} (1- y_p^{\xi_k} )]^T,
  \eeq
 for given $\xi_k$, where $\,y_p = -\ln \: (1-p)$. 
 
We assume that the true return level is composed of a model average of the return levels ($r_{k, true}$) from the submodels : 
\beq
 r_{true}\ =\ \sum_{k=1}^K w_k\; r_{k, true}.
 \eeq
 Let $\hat w_n(k)$ and $\hat r_n (k)$ represent the weight and return level estimates obtained from data with sample size $n$. Additionally, let $v_k$ denote the asymptotic variance of $\hat r_k$, as provided in (\ref{delta1}). Define $W_k$ as a random variable such that $\hat w_n(k) \xrightarrow{ \text{ D ~} } W_k$ as $n \rightarrow \infty$, where $E(W_k)= w_k$. Then we have the following result for the asymptotic  distribution of the MA return level estimator: \\
 
{\bf Theorem 1}: When submodel $M_k$ is fitted using the MLE, we have, as $n \rightarrow \infty$
\beq 
 \sum_{k=1}^K \hat w_n(k)\: \hat r_n (k) - r_{true}  \: \xrightarrow{ \text{ D ~} }\: \Lambda_{MA} \; =\; \sum_{k=1}^K w_k\: \Lambda_k 
 \eeq
 where $\Lambda_k\, \sim N (0,\: v_k)$. \\
 
{\bf Proof}:
Let $R_k$ be a normal random variable such that $\hat r_n (k) \xrightarrow{ \text{ D ~} } R_k \sim N (r_{k, true},\: v_k)$. Then, we have:
\beq \label{proof1}
\begin{aligned}
&\sum_{k=1}^K \hat w_n(k)\: \hat r_n (k) - r_{true}\ =\ \sum _{k=1}^K \{\hat w_n(k) -W_k)\; \hat r_n (k)
\; +\; \siK W_k \; \hat r_n (k) - r_{true} \\
& \ =\ \siK \{\hat w_n(k) - w_k\}\; \hat r_n (k)\; +\; \siK (w_k -W_k)\; \hat r_n (k)\; +\; \siK W_k\; \hat r_n (k) - r_{true} 
\end{aligned}
\eeq
 Since $\hat w_n(k)$ is uniformly integrable, it follows that $E[\hat w_n(k)] \rightarrow w_k $, which implies that $\hat w_n(k) \xrightarrow{ \text{ p ~} } w_k$ (Serfling 1980). Thus, the above equation simplifies to: 
 \beq \label{proof1-1}
 \begin{aligned}
& \ \xrightarrow{ \text{ D ~} }\ \siK  (w_k -W_k)\; R_k\; +\; \siK W_k R_k\; -\; \siK w_k\; r_{k, true}
\\
&\ =\ \siK w_k\; (R_k -r_{k, true})\ =\ \siK w_k\; \Lambda_k.
\end{aligned}
\eeq
The Slutsky theorem and the continuity theorem of weak convergence are applied in the above derivation. This establishes the statement of the theorem.\qed
\\

Unlike the individual $\Lambda_k$, the limiting random variable $\Lambda_{MA}$ is no longer normally distributed. The mean and variance of $\Lambda_{MA}$ are given as $E(\Lambda_{MA}) = 0 $ and
\beq \label{varLambda}
Var(\Lambda_{MA})\ =\  \uw^T\; C\; \uw,
\eeq
where  $\uw^T=(w_1,\dots,w_K)$ and $C$ is $Var(\underline{\Lambda})$ which is the asymptotic covariance matrix of $\hat \ur$, where $\underline{\Lambda}^T = (\Lambda_1, \dots, \Lambda_K)$.

When submodel $M_k$ is fitted using the LME, the mixed method of Morrison \& Smith (2002) and Ailliot et al.~(2011) or the penalized likelihood method of Coles \& Dixon (1999), the estimators also follow an asymptotic normal distribution with the corresponding covariance matrix. Therefore, Theorem 1 remains valid for these estimators as well.

\subsection{Asymptotic variance of MA return level}

The diagonal elements of $C$, the asymptotic variances of $\hat \ur$, are obtained using (\ref{delta1}). However, the off-diagonal elements of $C$,
representing the covariances $\text{Cov} (\hat{r}_i, \; \hat{r}_j)$ for $i \neq j$, are difficult to compute explicitly.
To address this, we considered a variance formula directly from (\ref{rlMA}):
\beq \label{varMA}
\text{Var} (\hat{r}_\text{MA}) = \sum_{k=1}^K w_k^2 \; \text{Var} (\hat{r}_k) + \sum_{i \neq j }^K w_i\, w_j \, \text{Cov} (\hat{r}_i, \; \hat{r}_j).
\eeq
 If the correlation coefficient ($\rho_{ij}$) between $r_i$ and $r_j$ is known, the second term on the right-hand side can be computed using the following formula: 
\beq \label{covij}
\text{Cov} (\hat{r}_j, \, \hat{r}_j) = \rho_{ij} \sqrt{\smash[b]{\text{Var}(\hat{r}_i)}} \,\sqrt{\smash[b]{\text{Var}(\hat{r}_j)}}.
\eeq
To approximate $\rho_{ij}$, we use the Pearson correlation coefficient between models $M_i$ and $M_j$, based on estimated values from both models. These estimated values include, for example, selected quantiles and parameter estimates. Specifically, we use nine quantile estimates ranging from 0.1 to 0.9 in increments of 0.1, along with the three-parameter estimates ($\hat\mu_i$, $\hat\sigma_i$, and $\xi_i$). These 12 values are treated as 12 observations for approximating the correlation. 

\subsection{MA variance with random weights}\label{sec:asvar}

For computing $Var(\Lambda_{MA})$ in real applications with data of size $n$, we may need to replace $\uw$ in (\ref{varLambda}) with its estimate $\hat \uw^T= (\hat w_1, \dots, \hat w_K)$. In this case, $\uwh$ is a random vector derived from the sample, meaning that the formula (\ref{varLambda}) may need to be adjusted to account for the uncertainty in $\uwh$. To calculate $Var(\Lambda_{MA})$ using the random vector $\uwh$, we assume that $\uwh$ and $\hat \ur$ are independent. The expression for the variance is then given by:
\beq \label{randwt}
\begin{aligned}
	Var (\hat{r}_\text{MA})\ &=\ Var (\uwh^T \hat \ur ) , ~~ \text{where} ~ \hat \ur^T =(\hat r_1, \dots, \hat r_K) \\
	\ &=\ Var_{\uwh} \left[  E (\uwh^T \hat \ur \, |\, \uwh ) \right]\; +\; E_{\uwh} \left[ Var ( \uwh^T \hat \ur \, |\, \uwh )\right] \\
\	& =\ Var_{\uwh} \left[ \uwh^T E(\hat \ur) \right]\; +\; E_{\uwh} \left[ \uwh^T\, C\, \uwh \right] \\
\	& =\ E(\hat \ur)^T\, D\, E(\hat \ur)\; +\; tr (D\, C)\; +\; E(\uwh)^T\, C\, E(\uwh),
\end{aligned}
\eeq
where $D= Var (\uwh)$.
In the last equation, we used the formula $E(\uwh^T\, C\, \uwh) = tr (D\, C) + E(\uwh)^T\, C \, E(\uwh)$ (Ravishanker et al.~2022, for example). Here, $C$ is approximated by the asymptotic covariance matrix $\hat C$, computed using (\ref{delta1}) and (\ref{covij}), and evaluated at $\hat \ur$. Due to the additional uncertainty introduced by the randomness of the weights, this variance must be greater than or equal to the asymptotic variance in (\ref{varMA}) for fixed weights.

We need to estimate $E(\uwh)$, $E(\hat \ur)$, and $D$ from the data for practical purposes. However, deriving these quantities theoretically from (\ref{weight}) 
is challenging. 
 One possible approach is to use bootstrapping to approximate these quantities, although we did not adopt this method in our study. Instead, we assume a distribution for $\uwh$. Given that $0 \le \hat w_k \le 1 $ for $k=1,\dots, K$ and that the weights satisfy $\sum_{k=1}^K \hat w_k =1$, it is reasonable to assume that 
 $\uwh$ follows a Dirichlet distribution with parameter vector $\uw$, where $0 \le w_k \le 1 $ for $k=1,\dots, K$ and $\sum_{k=1}^K w_k =1$. Then, under this assumption,
 \beq \label{Dirichlet}
 E(\hat w_k)= w_k,~~~ \text{Var}(\hat w_k)= w_k\, (1-w_k)/2, ~~~ \text{Cov}(\hat w_i, \hat w_j)= - w_i\, w_j /2
 \eeq
  for all $i \ne j$. These quantities form the elements of $D$. The covariance matrix $D$ can then be approximated by substituting $\hat \uw$ in place of $\uw$, yielding the estimator $\hat D$. 
 
 To approximate $E(\hat \ur)$, we used a moving average of order $q$ for $\hat \ur$ instead of directly substituting $\hat \ur$ into $E(\hat \ur)$.  
 With these approximations, we obtain $Var (\hat{r}_\text{MA})$ in (\ref{randwt}) using the following formulation:
\beq \label{est-randwt} 
\widehat{Var} (\uwh^T \hat \ur)\ \approx \  \widetilde{\hat \ur(q)}^T\, \hat D \; \widetilde{\hat \ur(q)}\;  +\; tr (\hat D\, \hat C)\; +\; \uwh^T\, \hat C\, \uwh . 
\eeq
where $\widetilde{\hat \ur (q)}^T$ is the order $q$ moving average of $\hat \ur^T=(\hat r_1, \dots, \hat r_K)$. We chose $q=3$ in this study. 
This formula is computationally much faster than using the bootstrap to estimate $E(\uwh)$, $E(\hat \ur)$, and $D$. 

\section{Simulation study}
\label{sec.simul}

\subsection{Computing strategy}

Selecting the number ($K$) of candidate models is an important step in the proposed method.
One recommended strategy is to start with a relatively large $K$ (e.g., $K = 18$) and then remove submodels with small weights, such as those with $w_k \le 0.01$.  Additionally, if the weight assigned to either the smallest $\xi_k$ or the largest $\xi_k$ remains above a certain threshold (e.g., 0.1), the method adds a few additional $\xi_k$ values beyond the lower or upper bounds. The weights are then recalculated using the updated number of submodels, denoted by $K^\prime$, and the MA return level estimate $\hat r_{MA}$ is obtained using the new weight vector ${\uwh^\prime}^T =(\hat w^\prime_{i_1},\dots,\hat w^\prime_{i_{K^\prime}})$. Thus, choosing the number $K$ may not be so critical provided that $K$ is not too small. We would recommend $K$ to be greater than or equal to seven, according to our experience. {Nevertheless, in Section \ref{sec:further}, we consider a systematic way to select an appropraite $K$.}

Thus, selecting between 90\% and 95\% confidence intervals for choosing $\xi_k$ values may no longer be critically important, as we adjust $\xi_k$s flexibly based on the assigned weights. For the same reason, selecting the initial method—whether using the profile likelihood in the MLE or bootstrap samples in the LME—to select $\xi_k$ is also not crucial. A pilot simulation study in the Supplementary Material indicates that variations in the starting method (`mle' or `lme') do not produce substantially different results. 
Similarly, selecting different confidence levels (90\% or 95\%) does not lead to substantial performance disparities. Therefore, in this study, we use the 95\% confidence interval based on the profile likelihood (`mle' starter) of $\xi$ to select $\xi_k$. 

 Additionally, we considered a `conventional' model averaging approach in which the parameters of each $GEV_k$ submodel are estimated using the MLE, and submodels are weighted based on smooth AIC scores. Here, the smooth AIC score (Claeskens \& Hjort 2008) is defined as $\Delta AIC_k = AIC_k - min_{1 \le k \le K} AIC_k$, so that the weight is obtained by:
 \beq \label{wtlh-2}
 \hat w_k\ =\ {\frac{ \exp(-\Delta AIC_k/2) }{\sum_{j=1}^K \exp(-\Delta AIC_j/2)}}.
 \eeq
 
 For comparative purposes, we consider two existing modified MLE methods. The mixed method proposed by Morrison \& Smith (2002) and Ailliot et al.~(2011) has at least two versions: (1) Re.MLE1 – This approach estimates the parameters by maximizing the likelihood under the constraint that the population mean ($\lambda_1$) is equal to the sample mean ($l_1$), and (2) Re.MLE2 - This version is obtained by maximizing the likelihood under the additional restrictions that
 \beq \label{cons2}
 \lambda_1\, =\, l_1, ~~~ \lambda_2\, =\, l_2,
 \eeq
 where $\lambda_2$ and $l_2$ represent the second population and sample L-moments, respectively. 
 
 The penalized likelihood method proposed by Coles and Dixon (1999) estimates the parameters by maximizing the following penalized log-likelihood:
 \begin{equation} \label{pnllk_gev}
 l_{pen}(\mu,\sigma,\xi)\; =\;  \ln (L(\mu,\sigma,\xi))\; +\; \ln ( p(\xi)) ,
 \end{equation}
 where $L(\mu,\sigma,\xi)$ is the likelihood function and $p(\xi)$ is a penalty function applied to $\xi$:
 \begin{equation}\label{p_coles}
 p(\xi)\ =\ \begin{cases} 1 & \mbox{ if $\xi \geq 0$}\\
 \exp\{-\lambda(\frac{1}{1+\xi}-1)^{\alpha}\} & \mbox{ if $-1< \xi < 0$}\\
 0 & \mbox{ if $\xi\leq -1$}\end{cases}
 \end{equation}
 with the hyperparameters $\alpha=1$ and $\lambda=1$.
 We refer to the estimator obtained using this method as `MLE.CD,' where CD stands for Coles and Dixon (1999). 
  
 For the LME, we utilized the `Lmoments' or `lmomco' packages in R, as provided by Karvanen (2006), based on the methods of Hosking (1990) and Asquith (2014). For the MLE, return level plots, and confidence intervals based on profile likelihood, we primarily used the `ismev' package (Coles 2001) in R. Additionally, for constrained optimization under (\ref{cons2}), we employed the `Rsolnp' package. 
 
  As presented in the Supplementary Material, a pilot simulation study revealed that MA methods exhibited substantial underestimation of return levels when $\xi \le -0.3$. We applied a trimming approach to address this issue by removing the smallest observations from the sample. 
   The abbreviations of the MA and other methods used in this study are listed in Table \ref{methods_MA}, where `trim=1' and `trim=2' indicate the cases where the minimum and second minimum observations, respectively, were excluded during estimation or weighting. Additionally, based on another pilot simulation study, we found that the weighting schemes `gLd' and `med' did not result in substantial performance differences. Therefore, for simplicity—given that a total of 12 methods are already considered—the MA method with `med' weighting is omitted from this simulation study.
 
  \begin{table}[htp]
 	\caption{Abbreviations of the model averaging and other methods used in this study. Here, `gLd' represents the generalized L-moments distance defined in (\ref{GL-dist}), the weighting formula using smooth AIC is provided in (\ref{wtlh-2}), the constraints for Re.MLE are defined in (\ref{cons2}), and the penalty function for MLE.CD is given in (\ref{p_coles}).} \label{methods_MA}
 	\vspace{.4cm}
 \centering
 \begin{tabular}{c| c| c}
 	\hline
 method name &estimation &weighting \\ \hline
  MA.gLd1  & MLE & gLd/trim=1 \\ \hline
  MA.gLd2  & MLE & gLd/trim=2 \\ \hline
  MA.like0 & LME & smooth AIC/trim=0  \\ \hline
  MA.like1 & LME & smooth AIC/trim=1  \\ \hline
  MA.cvt   & MLE & smooth AIC  \\ \hline
  Re.MLE1  & MLE/constraint-1 &  \\ \hline
  Re.MLE2  & MLE/constraint-2 &  \\ \hline
  MLE.CD   & MLE/penality-CD &  \\ \hline
 
\end{tabular}
\end{table}

\subsection{Simulation setting}

To evaluate the performance of the MA methodology, we conducted a simulation study  by generating random numbers from a GEVD in which high quantiles were known. 
 This study estimated the 100-year and 200-year return levels (corresponding to the 0.99 and 0.995 quantiles) using a sample size of $n=50$. The number of submodels ($K$) was set to 12, and the number of bootstrap samples was 500. We generated 1,000 random samples to compute the following evaluation measures for comparing different estimators:
 \beq \label{eval}
 \text{Bias} = \bar{\hat{r}} - r , ~~~~
 \text{SE} = \left\{\frac{1}{N} \sum_{i=1}^N (\hat{r}_i -\bar{\hat{r}})^2 \right\}^{1/2}, ~~~~
 \text{RMSE} =  \left\{ \frac{1}{N} \sum_{i=1}^N ({\hat{r}}_i - r)^2 \right\}^{1/2},
 \eeq
 where 
$r$ denotes the true return level, $\bar{\hat{r}} =  \sum_{i=1}^N \hat{r}_{i}/N$, $\hat{r}_{i}$ is the return level estimate from the $i$th sample, and $N$ represents the number of simulation samples ($N=1,000$). Lower values of SE and RMSE are preferable.

The upper endpoint of the GEVD is bounded when $\xi >0$ (under the Hosking-Wallis notation). As a result, estimating high quantiles of the GEVD when $\xi >0$ is relatively straightforward, and the RMSE values tend to be smaller. The MLE and LME typically perform well when $\xi \ge 0$. 
Therefore, we focus on cases when $\xi$ ranges over $(-0.5, 0.0)$. The case $\xi \le -0.5$ is not considered because the GEVD has no finite variance in this range. We conduct experiments for 10 values of $\xi$: $\xi = -0.45,\ -0.4,\ -0.35,\ -0.3,\ -0.25,\ -0.2,\ -0.15,\ -0.1,\ -0.05$, and $-0.001$. For better interpretability, we set $\mu=100$ and $\sigma=30$, as estimates of these parameters are equivariant (Casella \& Berger 2001). Setting $\mu=0$ and $\sigma=1$ often results in values less than one with decimal points, making it harder to distinguish differences among methods in the summary table. Additionally, this (0,1) scaling occasionally produces negative values, which in rare cases can cause numerical issues in computing the likelihood function or inverting the Hessian matrix.

\subsection{Simulation results}

\begin{table}[htp]
	\caption{Simulation results of the bias, standard error (SE), and root mean squared error (RMSE) of the GEVD, computed for the quantile $q_{0.99}$, using the ten estimation methods across ten values of $\xi$. The sample size is $n=50$, with parameters $\mu=100$ and $\sigma=30$. Acronyms of the method names are provided in Table \ref{methods_MA}.
	}\label{sim_result}
\vspace{.3cm}
	\centering
	\begin{tabular}{ll|rrrrrrrrrr}
		\hline
		Measure & Method & $-.45$ & $-.4$ & $-.35$ & $-.3$ & $-.25$ & $-.2$ & $-.15$ & $-.1$ & $-.05$ & $0 $ \\ \hline
		Bias & MA.gLd1 & -51.3 & -32.8 & -20.9 & -12.2 & -7.3 & -2.1 & 0.2 & -0.7 & -2.1 & 2.5  \\ 
		Bias & MA.gLd2 & -46.6 & -28.9 & -17.3 & -8.8 & -5.0 & 0.7 & 2.7 & 1.7 & 0.0 & 3.6 \\ 
		Bias & MA.like0 & -21.6 & -7.3 & -4.6 & -1.1 & 1.0 & 4.5 & 3.4 & 1.1 & -1.8 & 2.4  \\ 
		Bias & MA.like1 & -12.6 & 0.2 & 3.1 & 6.3 & 8.1 & 10.8 & 9.2 & 7.0 & 3.4 & 5.8  \\ 
		Bias & MA.cvt & 104.2 & 52.0 & 29.4 & 25.3 & 15.4 & 13.5 & 9.7 & 8.2 & 6.1 & 3.9 \\ 
		Bias & MLE & 56.0 & 45.0 & 28.5 & 23.4 & 13.1 & 13.3 & 8.0 & 2.1 & -3.1 & 0.4  \\ 
		Bias & MLE.CD & -53.2 & -31.9 & -20.9 & -14.1 & -10.9 & -5.6 & -5.9 & -7.6 & -9.7 & -6.5  \\ 
		Bias & Re.MLE1 & -4.7 & 6.3 & 6.4 & 7.2 & 5.1 & 6.9 & 2.8 & -1.6 & -7.0 & -4.4  \\ 
		Bias & Re.MLE2 & -21.2 & -5.8 & -2.1 & 1.6 & 4.0 & 7.1 & 5.4 & 2.4 & -0.8 & 3.2  \\ 
		Bias & LME & -15.4 & -4.4 & 2.2 & 4.3 & 4.4 & 7.0 & 4.0 & 0.4 & -4.0 & -0.6  \\ \hline
		
		SE & MA.gLd1 & 165.8 & 136.5 & 115.2 & 96.0 & 82.6 & 65.1 & 56.0 & 43.8 & 35.2 & 31.2 \\ 
		SE & MA.gLd2 & 166.5 & 136.7 & 115.3 & 96.3 & 82.8 & 65.4 & 56.5 & 44.1 & 35.5 & 31.5  \\ 
		SE & MA.like0 & 194.0 & 164.1 & 125.4 & 107.0 & 89.0 & 69.7 & 58.4 & 46.3 & 36.3 & 32.6  \\ 
		SE & MA.like1 & 194.2 & 164.6 & 126.7 & 108.1 & 90.9 & 71.2 & 60.0 & 47.4 & 37.1 & 33.1  \\ 
		SE & MA.cvt & 316.7 & 229.7 & 161.4 & 146.9 & 127.8 & 87.8 & 70.2 & 55.1 & 44.3 & 38.1  \\ 
		SE & MLE & 273.1 & 214.8 & 156.9 & 135.0 & 103.6 & 84.3 & 68.0 & 52.7 & 39.4 & 35.5  \\ 
		SE & MLE.CD & 164.0 & 135.3 & 114.8 & 97.6 & 82.5 & 69.6 & 60.6 & 49.6 & 41.6 & 39.0  \\ 
		SE & Re.MLE1 & 193.4 & 161.4 & 130.1 & 112.9 & 94.1 & 77.6 & 67.0 & 54.0 & 44.5 & 41.6  \\ 
		SE & Re.MLE2 & 190.3 & 165.1 & 127.6 & 108.9 & 90.1 & 70.8 & 59.8 & 47.4 & 37.3 & 33.6  \\ 
		SE & LME & 189.3 & 161.2 & 128.7 & 110.5 & 91.7 & 73.7 & 62.8 & 49.9 & 39.4 & 35.3  \\ \hline
		
		RMSE & MA.gLd1 & 173.6 & 140.4 & 117.1 &  96.8  & 82.9 & 65.1 & 56.0 & 43.8 & 35.3 &  31.3 \\
		RMSE & MA.gLd2 & 172.9 & 139.7 & 116.6 &   96.7 &   83.0 & 65.4 & 56.6 & 44.1 & 35.5 &  31.7 \\
		RMSE & MA.like0 & 195.2 & 164.3 & 125.5 & 107.0 &   89.0 & 69.8 & 58.5 & 46.3 & 36.3 &  32.7\\
		RMSE & MA.like1 & 194.6 & 164.6 & 126.7 & 108.3 &   91.3 & 72.0 & 60.7 & 47.9 & 37.3 &  33.6\\
		RMSE & MA.cvt & 333.4 & 235.5 & 164.1 & 149.1 & 128.7 & 88.8 & 70.9 & 55.7 & 44.7 &  38.3\\
		RMSE & MLE & 278.8 & 219.5 & 159.5 & 137.0 & 104.4 & 85.3 & 68.5 & 52.7 & 39.5 &  35.5\\
		RMSE & MLE.CD & 172.4 & 139.0 & 116.7 &   98.6 &   83.2 & 69.8 & 60.9 & 50.2 & 42.7 &  39.5\\
		RMSE & Re.MLE1 &  193.5 & 161.5 & 130.3 & 111.1 &   94.2 & 77.9 & 67.1 & 54.0 & 45.0 &  41.8\\
		RMSE & Re.MLE2 & 191.5 & 165.2 & 127.6 & 108.9 &   90.2 & 71.2 & 60.0 & 47.5 & 37.3 &  33.8\\
		RMSE & LME & 189.9 & 161.3 & 128.7 & 110.6 &   91.8 & 74.0 & 62.9 & 49.9 & 39.6 &  35.3\\
		\hline
	\end{tabular}
\end{table}

\begin{figure}[h!bt]
	\hspace{-1.5cm}
	\begin{tabular}{c}	\includegraphics[width=8.8cm, height=7.5cm]{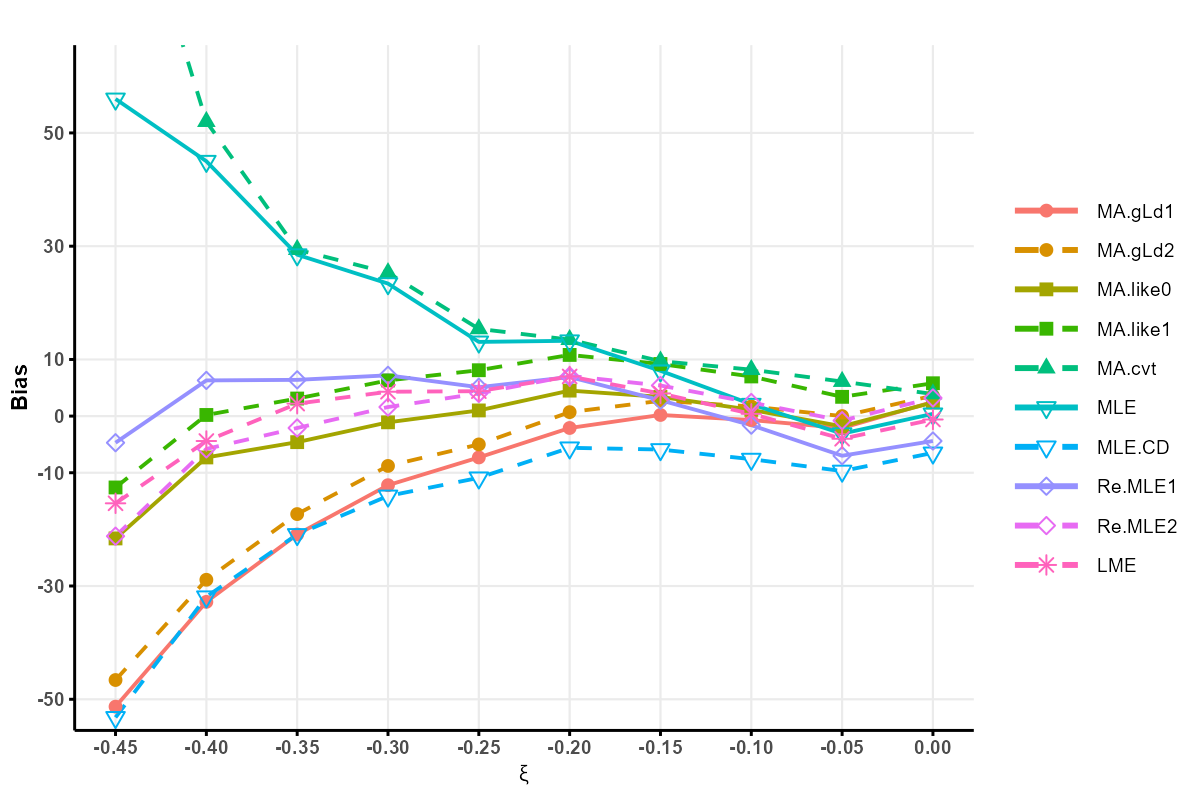}
		\includegraphics[width=8.8cm, height=7.5cm]{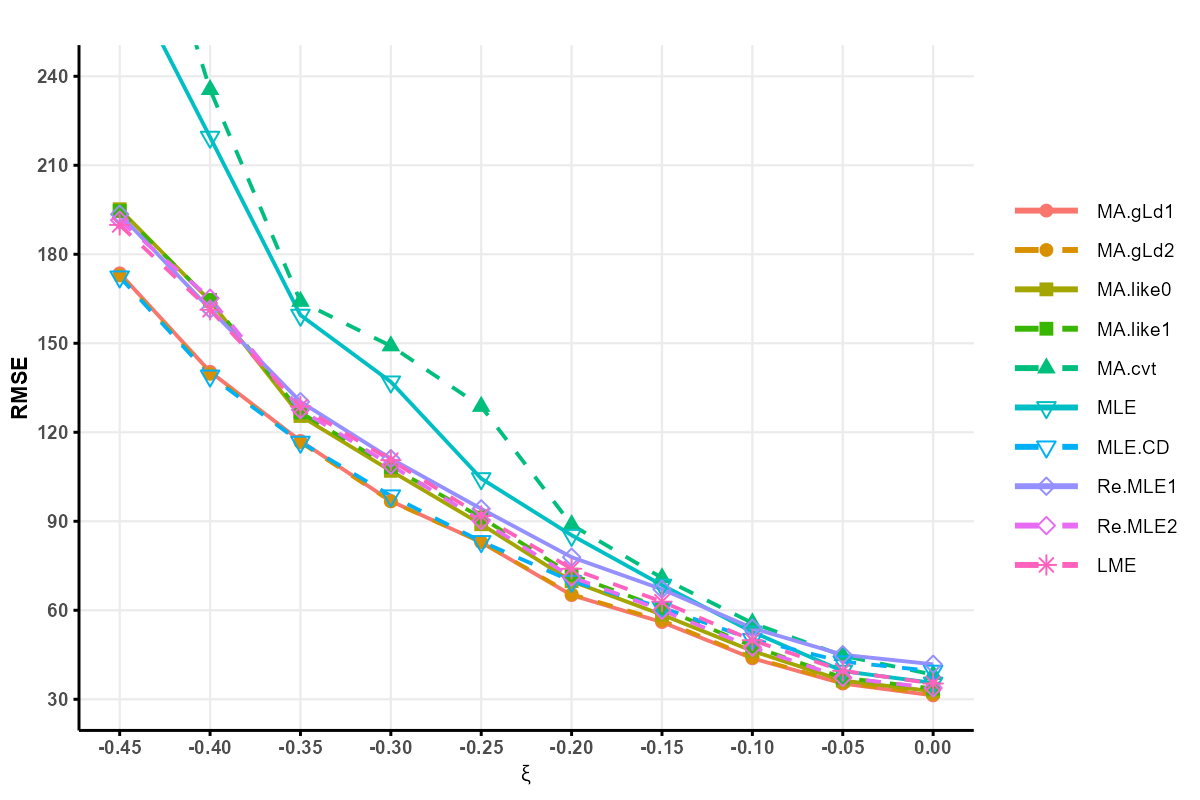}
	\end{tabular}
	\caption{{Same as Table \ref{sim_result} but for the bias (left) and RMSE (right).} 
	} \label{Hae-nam_biasrmse}
\end{figure}

Table \ref{sim_result} and Figure \ref{Hae-nam_biasrmse} present the simulation results for bias, SE, and RMSE computed for the quantile $q_{0.99}$ using ten estimation methods. Acronyms for the method names are provided in Table \ref{methods_MA}. Since the results for the `gLd' and `med' weighting schemes were similar in this simulation study, only the results for the `gLd' weighting scheme are included in Table \ref{sim_result}.

Regarding bias, the MLE exhibits overestimation (positive bias), particularly when $\xi \le -0.15$, while the proposed MA methods with `gLd' weight and MLE.CD display underestimation (negative bias), especially for $\xi \le -0.3$. The LME, Re.MLEs, and MA.likes generally perform well. For cases where $-0.25 < \xi < 0$ (moderately heavy-tailed distributions), the LME, MA.gLds, and MA.like0 perform well, producing small biases. For $-0.35 \le \xi \le -0.25$, MA.like0, Re.MLE2, and LME show good performance. For extremely heavy-tailed cases ($\xi < -0.35$), Re.MLE1 and MA.like1 perform well in terms of bias. 

 Regarding SE and RMSE, for $\xi \le -0.25$, MA.gLds and MLE.CD exhibit smaller SE values, resulting in lower RMSE values than other methods. However, this reduction in SEs is primarily due to severe underestimation, making the lower RMSE in these cases less meaningful. Nevertheless, it is noteworthy that MA.gLds remains both smallest in SE and bias for $-0.25 < \xi \le 0$, leading to genuinely lower and meaningful RMSEs. For $-0.35 \le \xi \le -0.25$, MA.likes, Re.MLE2, and LME yield smaller RMSE values than other methods. Determining the best-performing methods for $ \xi < -0.35$ is difficult, but LME, Re.MLEs, and MA.likes are acceptable choices. 
 
 Based on these findings, the most suitable method depends on the value of $\xi$: MA.gLds for $-0.25 < \xi \le 0$, MA.likes, Re.MLE2, and LME for $-0.35 \le \xi \le -0.25$, and MA.like1, Re.MLE1, and LME for $ \xi < -0.35$. 
 Considering bias and RMSE, MLE and MA.cvt do not perform well for the range of $\xi < 0$. Without specifying the $\xi$ value, Re.MLEs, LME, and MA.likes are generally reliable methods.

\section{Real data application}
\label{appl}

\begin{figure}[h!bt]
		\hspace{-1.5cm}
	\begin{tabular}{c}	\includegraphics[width=8.9cm, height=8.5cm]{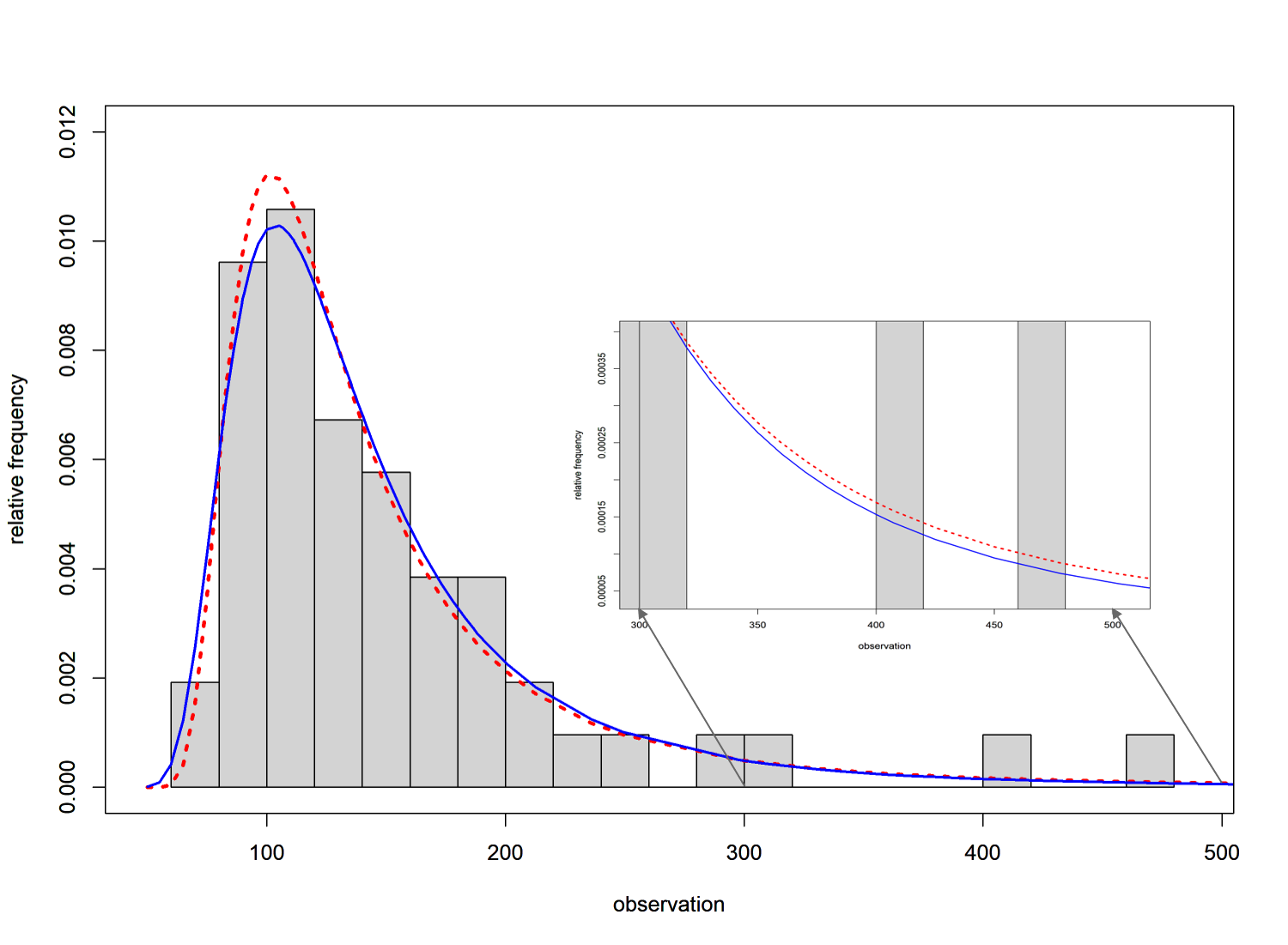}
		\includegraphics[width=9.1cm, height=9.1cm]{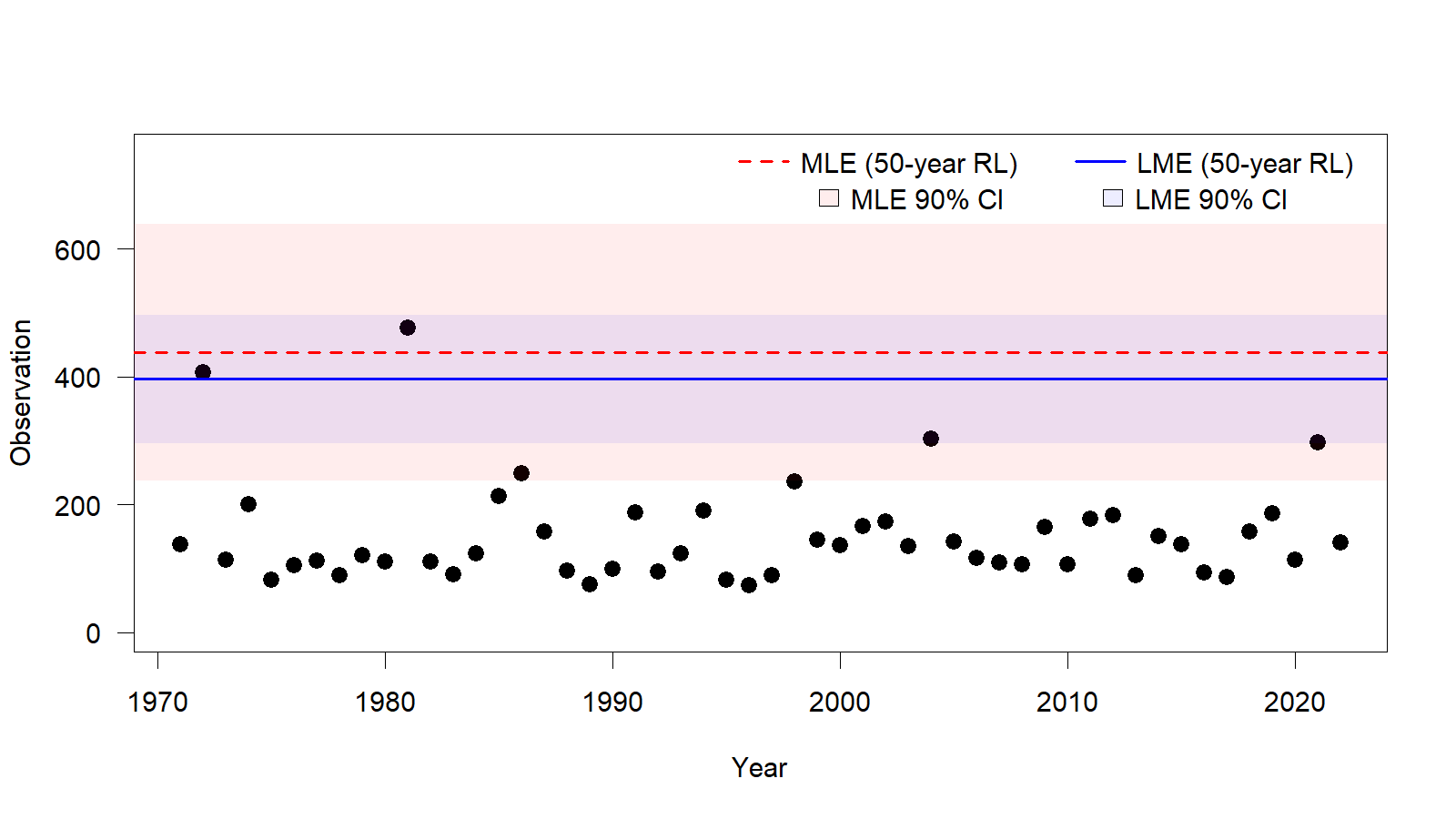}
	\end{tabular}
	\caption{(Left) Density plots overlaid on the relative frequency histogram and (Right) a scatterplot of observations by year with horizontal lines representing the 50-year return level estimates together with shaded regions for 90\% confidence intervals. The data depict the annual maximum daily rainfall (unit: $mm$) in the Hae-nam station, Korea. In both figures, the blue solid line and the red dashed line correspond to the estimates obtained using the LME and MLE methods, respectively.} \label{Hae-nam_hist}
\end{figure}

For a real-data example, we analyzed a time series of annual maximum daily precipitation (unit: $mm$) recorded in Hae-nam, Korea. The dataset consists of 52 observations from 1971 to 2022 and is available at the GitHub repository https://github.com/yire-shin/MA-gev/tree/main/data.
 We selected this weather station because the data exhibit a heavy right tail, making it well-suited for evaluating the effectiveness of the proposed methods. Figure \ref{Hae-nam_hist} displays the relative frequency histogram with density curves and a scatterplot with lines representing the 50-year return level estimates together with shaded regions for 90\% confidence intervals.

Table~\ref{Hae-nam_result} presents the parameter estimates, 100-year return levels, and their SEs obtained using the formula (\ref{randwt}) and the nonparametric bootstrap (NpB) with 500 resamples for 10 estimation methods. The number of submodels ($K$) was set to 12. Parameter estimates for the MA methods are derived from the surrogate models. In Table~\ref{Hae-nam_result}, the asymptotic SEs of MA.gLds (MA.likes) obtained using the formula (\ref{randwt}) are greater (less) than the corresponding SEs computed via nonparametric bootstrap. The SE (NpB) of MA methods is smaller than that of other methods, whereas MLE, MLE.CD, and Re.MLE1 yield larger SEs. The SEs of Re.MLE2 and LME fall within the moderate range—the 100-year return levels estimated by MA.likes and Re.MLE2 lie between those obtained using LME and MLE. Based on the findings from the simulation study, the return levels of MA.gLds may be underestimated despite their small SE (NpB) values. Conversely, the return level estimated by MLE may be overestimated, with exceptionally large SEs in both asymptotic and bootstrap calculations.

\begin{table}[ht]
	\caption{Estimates of parameters ($\hat \mu$, $\hat \sigma$, $\hat \xi$), 100-year return level (100-y RL), and their asymptotic standard errors (Asym.~SE) computed using the formula (\ref{randwt}) and nonparametric bootstrap (NpB) for 10 estimation methods applied to the annual maximum rainfall data  (unit:~$mm$) in Hae-nam station, Korea. 	}  \label{Hae-nam_result}
	\vspace{.5cm} 
	\centering 
		\begin{tabular}{c|cccccc}
			\hline
			Method& $\hat \mu$ & $\hat \sigma$ &$\hat \xi$ & 100-y RL & Asym.~SE &  SE (NpB)  \\  \hline
			MA.gLd1 & 115.3& 34.34& -0.336 & 492.2 & 73.0 & 66.3 \\
			MA.gLd2 & 115.1& 34.35& -0.342 & 498.5 & 74.1 & 67.6 \\ 
			MA.like0 & 113.4 & 34.72 & -0.351 & 511.5 & 67.3 & 92.3 \\
			MA.like1 & 114.8 & 33.94 & -0.363 & 518.1 & 72.1 & 93.9 \\ 
			MLE  &  112.6 & 35.10 & -0.394 & 569.4 & 204.8 & 211.8 \\
			MLE.CD & 113.3 & 35.23 &  -0.348 & 513.5 & & 130.2 \\
			Re.MLE1 & 111.5 &  33.88 &  -0.382 & 537.1 & & 133.2 \\
			Re.MLE2 & 112.8 &  34.58 &  -0.356 & 515.7 & & 103.8\\
			LME  &  113.5 & 37.35 & -0.310 &  494.9 & & 96.1 \\ \hline
		\end{tabular}
	\end{table}	
				
\begin{figure}[h!bt]
	\centering
	\begin{tabular}{c}	\includegraphics[width=13cm, height=11cm]{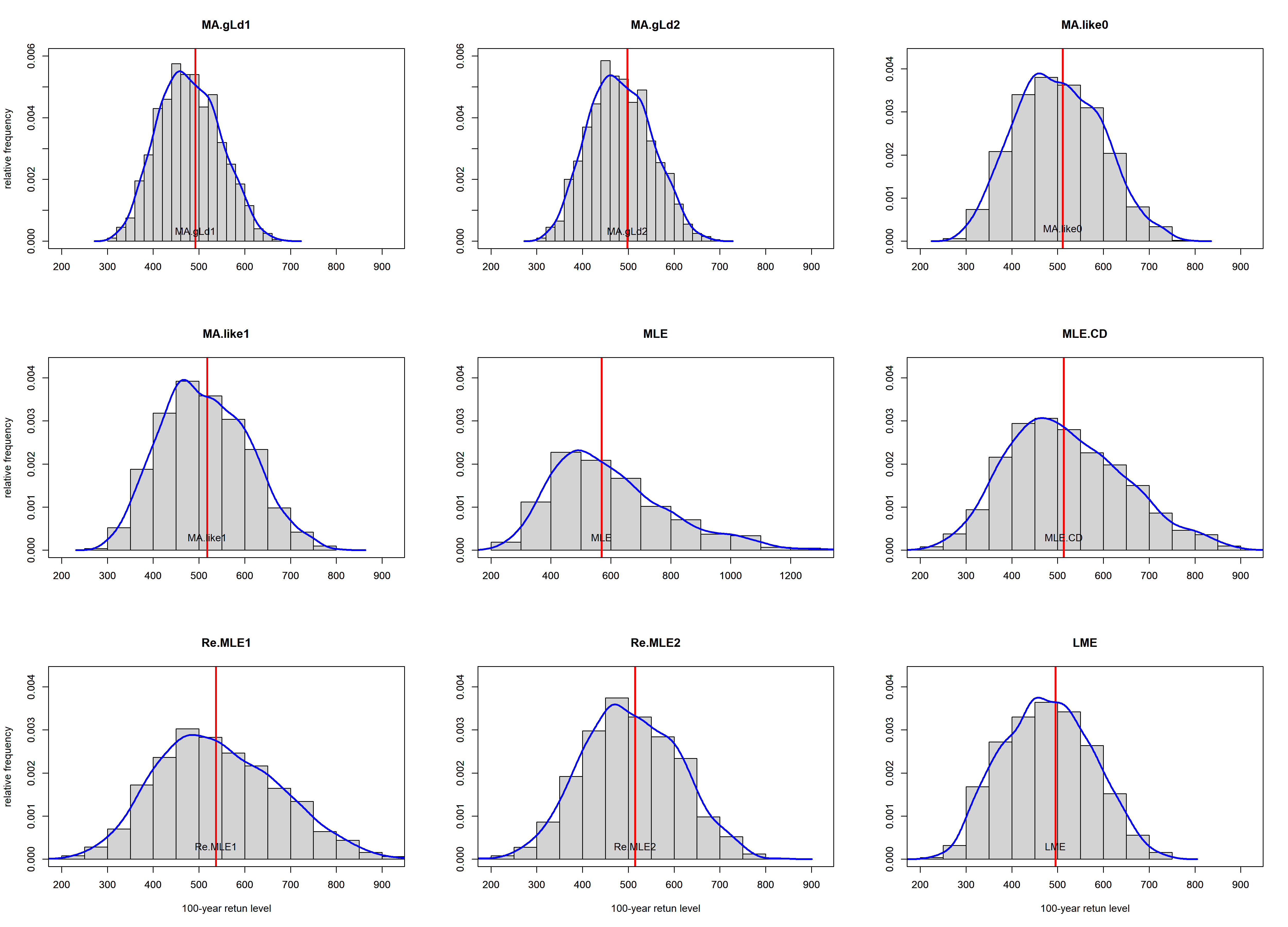}
	\end{tabular}
	\caption{Sampling distributions of the 100-year return levels from nine estimation methods, displayed as density plots overlaid on the relative frequency histogram. These distributions were obtained using 1,000 bootstrap samples from the annual maximum daily rainfall data (unit: $mm$) in Hae-nam, Korea. Acronyms for the method names are provided in Table \ref{methods_MA}.} \label{Hae-nam_samhist}
\end{figure}

Figure \ref{Hae-nam_samhist} presents the sampling distributions of the 100-year return levels from nine estimation methods, displayed as density plots overlaid on the relative frequency histogram, based on 1,000 bootstrap samples. Except for the MLE, all distributions appear approximately normal, though some exhibit slight right skewness.

\begin{figure}[h!bt]
			\centering
			\begin{tabular}{c}	\includegraphics[width=13cm, height=11cm]{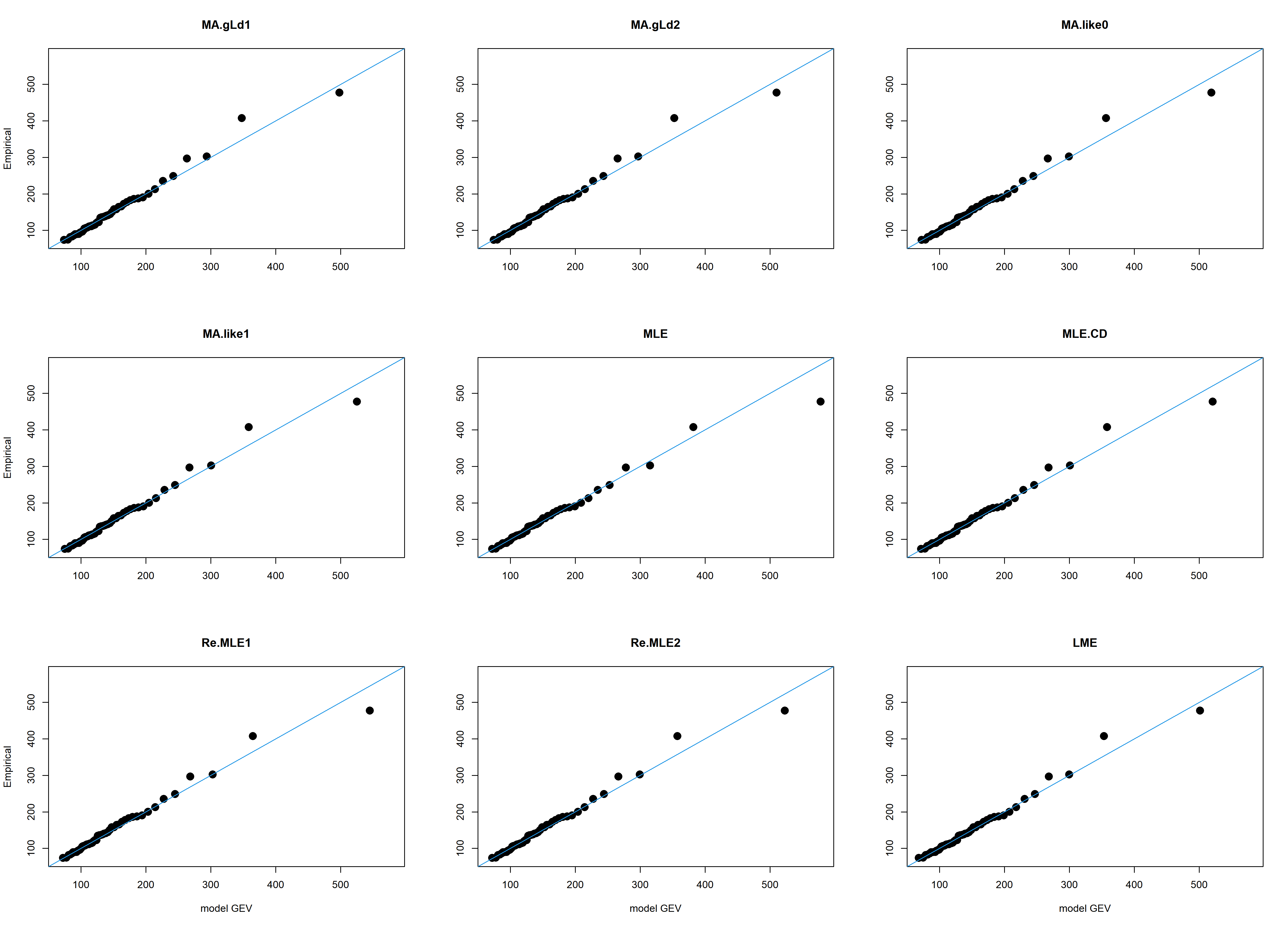}
			\end{tabular}
			\caption{Quantile-per-quantile plots of nine methods fitted to the annual maximum daily rainfall (unit: $mm$) data in Hae-nam, Korea. } \label{Hae-nam_qq}
		\end{figure}

Figure \ref{Hae-nam_qq} presents quantile-per-quantile plots for nine estimation methods. We used the standard plotting position formula: $q_i = (i -0.5)/n$ to construct these plots. Surrogate models were utilized for the MA methods. The plots forthe  MLE and Re.MLE1 appear to overestimate the largest observation, whereas the other methods fit the data relatively well. 
										
\begin{figure}[h!bt]
	\centering
	\begin{tabular}{c}	\includegraphics[width=13cm, height=11cm]{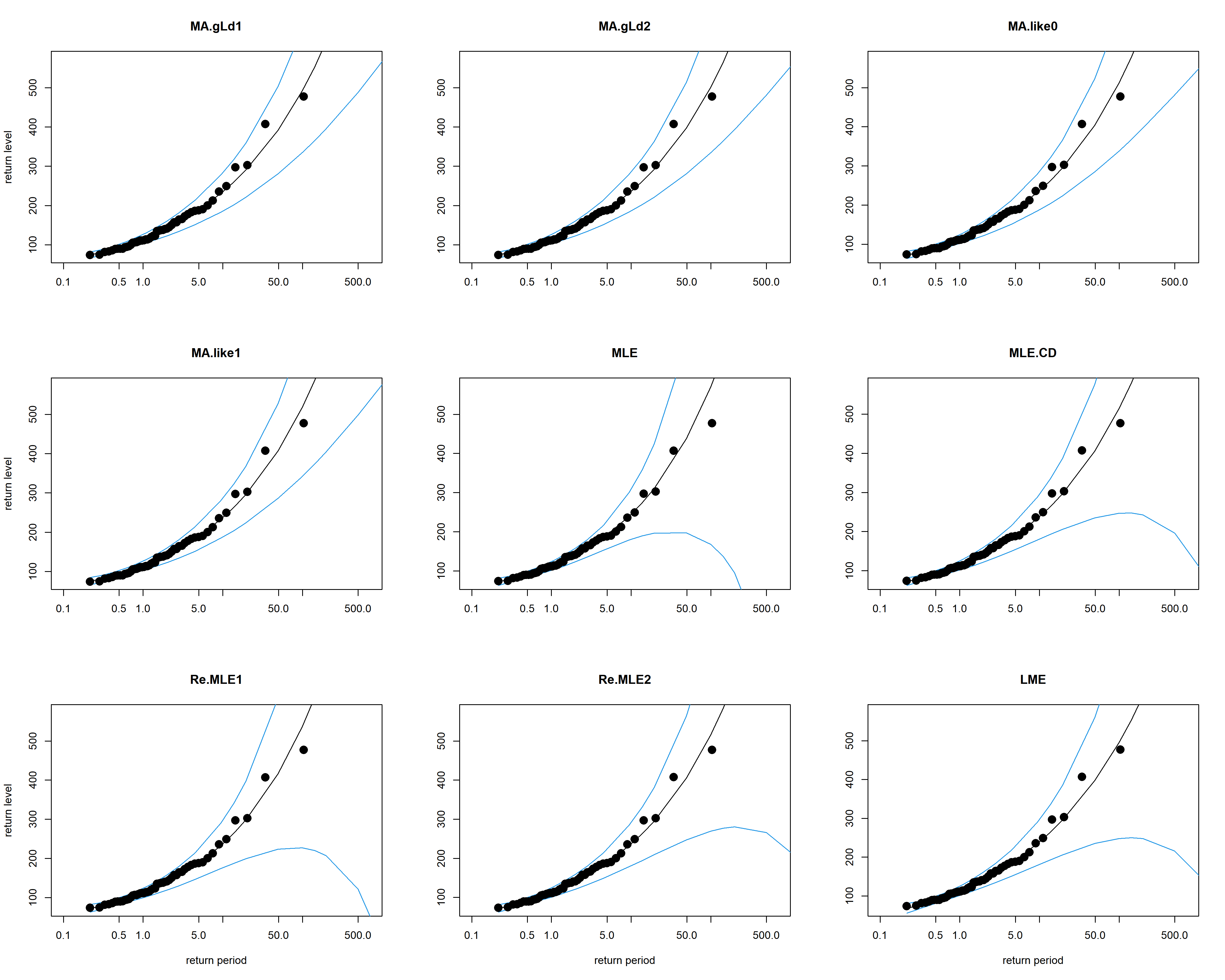}
	\end{tabular}
	\caption{Return level plots for nine estimation methods applied to the annual maximum daily rainfall data (unit: $mm$) from Hae-nam, Korea.
	} \label{Hae-nam_rlplot}
\end{figure}

Figure \ref{Hae-nam_rlplot} presents the return level plots for nine estimation methods. This figure was generated using the `gev.rl' function from the `ismev' package in R. To compute the covariance matrix of parameter estimates required for `gev.rl', we employed a nonparametric bootstrap with 500 repetitions. Because of the simplicity of interpretation, and because the choice of scale compresses the tail of the distribution so that the effect of extrapolation is highlighted, return level plots are particularly convenient for both model presentation and validation (Coles 2001). In Figure \ref{Hae-nam_rlplot}, the MA methods exhibit the narrowest confidence bands, while Re.MLE2 and LME show the second-best performance among the nine methods.

\section{Additional study} \label{sec:further}
\subsection{Selection of the number of submodels ($K$)} \label{sec:selK}

{In this section, we propose a systematic way to select an appropriate $K$. If $K$ is too large, unnecessary computation is required; if $K$ is too small, the variation of MA estimators due to the use of different $K$ values can be high. A sensitivity analysis by varying $K$ from 4 to 20 provides information on the performance of the estimation method with $K$. We propose a method considering two measures simultaneously. The first measure is the stability of the (100-year) return level estimates as $K$ varies from 4 to 20. The second measure is the standard error of the return level estimate. }	
	
\begin{figure}[h!bt]
 	\centering
 	\begin{tabular}{c}	\includegraphics[width=13cm, height=10cm]{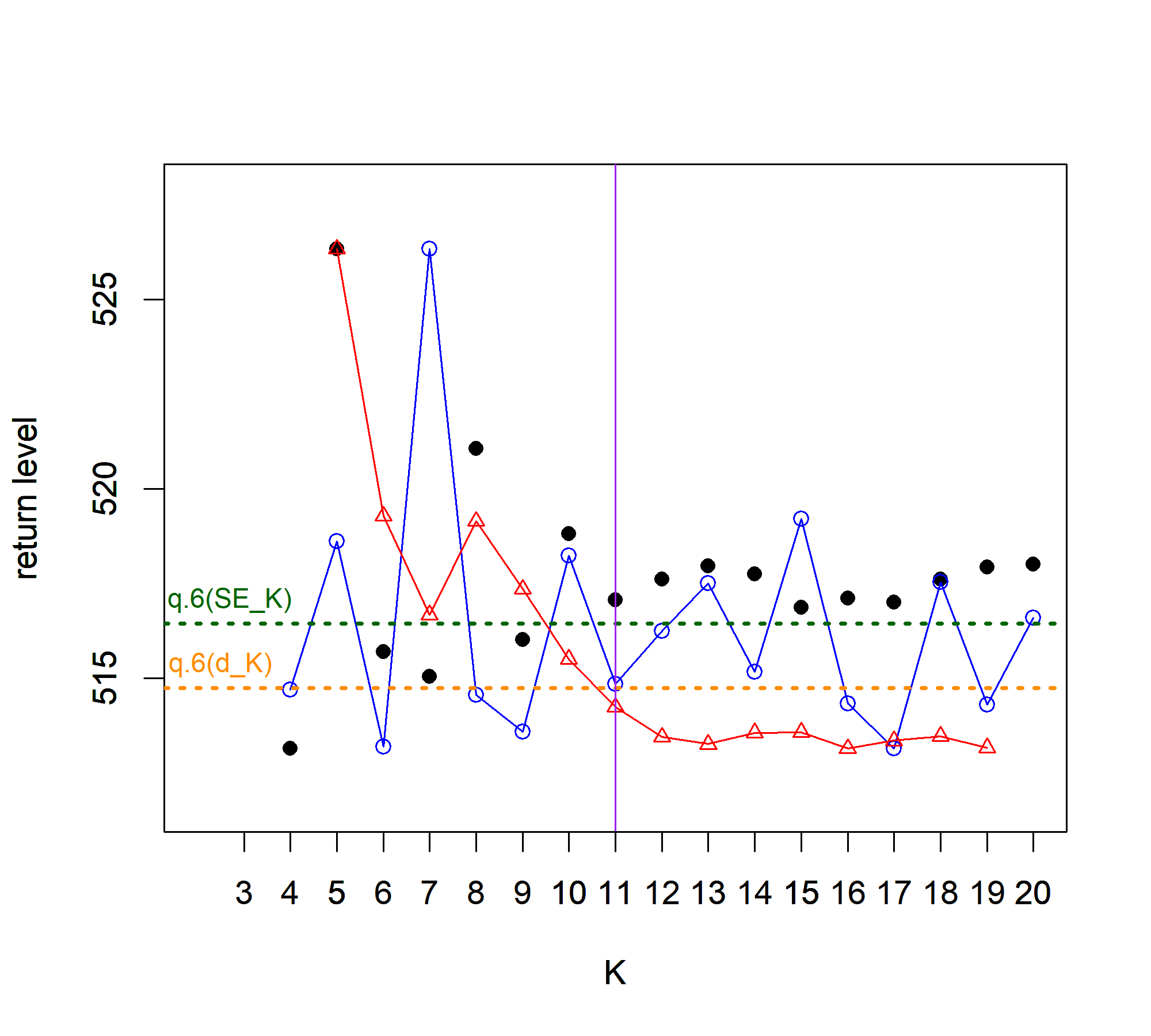}
 	\end{tabular}
 	\caption{A sensitivity analysis to select the optimal $K$, applied to the annual maximum daily rainfall data (unit: $mm$) from Hae-nam, Korea. The black dots indicate the 100-year return level estimates obtained by the MA.like1 method, the red solid line with triangular markers indicates the stability measures defined in (\ref{stability}), and the blue solid line with circular markers depicts the standard errors of return level estimates.
 	} \label{fig:selK}
 \end{figure}

{We define the stability measure of the return level estimate with respect to $K$ as
	\beq \label{stability}	
	d_K= |\hat{r}_\text{MA} (K)\, -\, \hat{r}_\text{MA} (K-1)| + 
          	|\hat{r}_\text{MA} (K+1)\, -\, \hat{r}_\text{MA} (K)|, 	
	\eeq
 for $K = 5, \dots, 19$. 
 The smaller $d_K$ means that $\hat{r}_\text{MA}(K)$ is more stable. 
 Figure \ref{fig:selK} shows a sensitivity analysis by varying $K$ from 4 to 20, with the MA.like1 method for the Hae-nam data. The black dots indicate the 100-year $\hat{r}_\text{MA} (K)$. 
 The red solid line with triangular markers indicates the stability measures $d_K$ rescaled to the range of $\hat{r}_\text{MA} (K)$. We treat the estimate as stable when $d_K$ is less than or equal to the $\alpha$-quantile of $d_K,\, K=5,\dots,19$. We employed $\alpha=0.6$ in this study. The $\alpha$-quantile is denoted by $q_{\alpha}(d_K)$. The orange horizontal dashed line stands for $q_{.6}(d_K)$ of rescaled $d_K$ values. In Figure \ref{fig:selK}, the estimates for $K \le 10$ are unstable in terms of $d_K$. Whereas, the estimates for $K \ge 11$ are stable. We will choose an appropriate $K$ among these values which lead to stable estimates. Let us denote a set of such $K$s by $I_S = \{ K: d_K \le q_{\alpha}(d_K) \}$. }
 
 {For the second measure, we want to select the $K$ such that the standard error (SE) of $\hat{r}_\text{MA} (K)$ is relatively small. The square root of the formula (\ref{est-randwt}) is employed for computing the SE, which is denoted by $SE_K$. We will choose an appropriate $K$ for which $SE_K$ is less than or equal to the $\alpha$-quantile of these SEs. We employed $\alpha=0.6$ in this study. The $\alpha$-quantile is denoted by $q_{\alpha}(SE_K)$. The blue solid line with circular markers in Figure \ref{fig:selK} depicts the $SE_K$ values rescaled to the range of $\hat{r}_\text{MA} (K)$. The green dashed horizontal line stands for $q_{.6}(SE_K)$ of rescaled $SE_K$ values.}
 
 {Let $I_E = \{ K: SE_K \le q_{\alpha}(SE_K) \}$. The $K^\prime$ is selected as the minimum among those $K$s that satisfy the stability and SE conditions simultaneously. In other words, 
 	\beq \label{Kprime}
 	K^\prime = \text{min}\, \{K:\, K \in I_S \cap I_E \}. 	
 	\eeq
 	Then the optimal $K^*$ is selected as the $K$ which leads to a smaller standard error near $K^\prime$, for example, among $K^\prime_N = (K^\prime-1,\, K^\prime-2,\, K^\prime,\, K^\prime+1,\, K^\prime+2$) which satisfy the stability condition.}
 	{That is, 
 	\beq \label{Kstar}
 	K^* = \text{arg}\, \text{min}\, \{ SE_K\} ~~ \text{for}~~ K \in K^\prime_N \cap I_S .
 	\eeq
 	In Figure \ref{fig:selK}, $K^\prime = 11$ which coincides with $K^*$. The vertical line represents the optimal $K^*$. It is notable that $K=17$ has the minimum standard error among stable estimates, but it may be too large and the estimates for $K=11$ and $K=17$ look similar. Thus we chose $K^*=11$ as the optimal number of submodels.
 Another example of choosing the optimal $K$ in this way is presented in the Supplementary Material.}

\subsection{Bias reduction and Bayesian model averaging}

The proposed MA methods with `gLd' and `med' weights exhibit negative biases despite the left-trimming approach adopted in this study, mainly when $\xi$ is less than $-0.3$. 
To address this issue, a bias correction technique may be necessary. Bayesian model averaging (BMA) could provide a solution for correcting such biases. 
 BMA assigns weights proportional to the model’s marginal likelihood (Hoeting et al.~1999; Raftery et al.~2005; Shin et al.~2019; Vettori et al.~2020, for example among many others). The BMA framework begins by specifying prior probabilities $p(M_k)$ for all candidate models $M_1,\dots, M_K$ and defining the prior density $p(\ux | M_k)$. The distribution of the return level given the data ($\ux$) is then expressed as:
\beq \label{bma-1}
\begin{aligned}
	p (r | \ux) \ = \int p(r, M_k | \ux)\, dM_k \	 \approx \ \sum_{k=1}^K p(r | M_k, \ux) \ p(M_k | \ux).
\end{aligned}
\eeq
Here, $p(M_k | \ux)$ can be expressed using Bayes' theorem as follows:
\beq \label{Bayes}
p(M_k | \ux)\ =\  \frac{ p(\ux | M_k)\; p(M_k) }{\sum_{k=1}^K p(\ux | M_k)\; p(M_k) },
\eeq
where $p(\ux | M_k)$ is typically the likelihood of model $M_k$, which corresponds to  (\ref{probMN}) for MA.gLd method and $p(M_k)$ represents the prior probability of $M_k$.
The posterior mean and variance of the BMA-predicted return level are given by (Hoeting et al.~1999; Claeskens \& Hjort 2008)
\beq \label{bma-mean}
\begin{aligned}
	E(r | \ux)\ &=\ \sum_{k=1}^K E (r |  M_k, \ux) \: w_k , \\
	Var (r | \ux)\ &=\ \sum_{k=1}^K \left[  E (r |  M_k, \ux) - E(r | \ux)  \right]^2 \: w_k \
	+\ \sum_{k=1}^K Var (r | \ux, M_k) \: w_k,
\end{aligned}
\eeq
where the weight $w_k$ is the same as (\ref{Bayes}). For estimating the return level in the GEV, we estimate $E (r | M_k, \ux)$ by $r( \hat \mu(\xi_k), \hat \sigma(\xi_k), \xi_k )$ in this study. Then the estimator of $E(r | \ux)$ in (\ref{bma-mean}) is the same as $\hat {r}_\text{MA}$ in (\ref{rlMA}), but with different weights. The weights in (\ref{weight}) and (\ref{wtlh-2}) do not use the prior $p(M_k)$, which is different from the weights in (\ref{bma-mean}).
The first term on the right hand side of variance formula (\ref{bma-mean}) is the among-model variance, and the second term is the within-model variance. In this study, $Var (r | \ux, M_k)$ is calculated using the asymptotic variance formula (\ref{delta1}).

{The selection of a prior is important to reduce the bias. We consider the prior for the shape parameter $\xi$ only, not for the location and scale parameters, because $\xi$ is unit-free and affects return level estimation sensitively. Indeed, some authors such as Coles and Dixon (1999), Martins and Stedinger (2000), and Cannon (2010) considered a prior or a penalty function for $\xi$ only. Whereas, Lee et al.~(2017) and Shin et al.~(2025b) suggested an approach using the data-adaptive penalty function (PF) which provides different hyperparameters according to an earlier estimate (LME in this study) of $\xi$. The approach using the data-adaptive PF turned out to be effective in correcting the estimation bias and robust to a poor selection of the prior (Lee et al.~(2017); Shin et al.~(2025b)). The data-adaptive PF may be viewed as an empirical prior.}

{We considered a PF based on the normal density:
\begin{eqnarray} \label{pen.norm}
p(M_k)\ =\ p(\xi_k)\ =  \psi (\xi_k;\: \mu_\xi, \sigma_\xi)
\end{eqnarray}
where $\psi (\xi_k;\: \mu_\xi, \sigma_\xi)$ represents a normal density of $\xi_k$ with a mean $\mu_\xi$ and standard deviation $\sigma_\xi$. The behavior of this PF depends on the choice of hyperparameters, $\mu_\xi$ and $\sigma_\xi$.  We chose the following hyperparameters, varying by the weighting method: For MA.gLd or MA.med,}
\beq \label{normalPF.gld}
\begin{aligned}
&\mu_\xi = \begin{cases} 1.5 \times \hat \xi &~~\text{if}~ \hat\xi > -0.45 \\
	                    1.5 \times (-0.45)  &~~\text{if}~ \hat\xi \le -0.45 \end{cases} \\
&\sigma_\xi = \begin{cases} (0.4 + \hat \xi)/4 + 0.14 &~~\text{if}~ \hat\xi > -0.4 \\
	    0.14                      &~~\text{if}~ \hat\xi \le -0.4, \end{cases} 
  \end{aligned}
\eeq
where $\hat \xi$ is the LME of $\xi$. For the MA.like method,
  \beq \label{normalPF.like}
  \begin{aligned}
&\mu_\xi = \begin{cases} 2.2 \times \hat \xi &~~\text{if}~ \hat\xi > -0.5 \\
	2.2 \times (-0.5)  &~~\text{if}~ \hat\xi \le -0.5 \end{cases} \\
&\sigma_\xi = \begin{cases} (0.45 + \hat \xi)/5 +0.11 &~~\text{if}~ \hat\xi > -0.45 \\
	0.11                       &~~\text{if}~ \hat\xi \le -0.45. \end{cases} 
\end{aligned}
\eeq
{The above PF is dependent on $\hat \xi$ (LME of $\xi$), because the underestimation bias of MA estimate is larger (smaller) for more (less) negative values of $\xi$. 
By providing more weight for $\xi_k$ closer to $-1.0$, we expect the MA method to correct the underestimation bias of $\hat {r}_\text{MA}$. When $\hat \xi \ge 0$, we set $p(\xi_k) =1/K$ for all $k$.
Some (five) constants in the above PFs were acquired by tuning the PF so as to approximately minimize the biases, using Monte Carlo simulations several times. One can of course use different tuning constants or different PFs. See Shin et al.~(2025b) for a data-adaptive beta PF.}

\begin{figure}[!htb]
	\centering 
	\includegraphics[width=14cm, height=11cm]{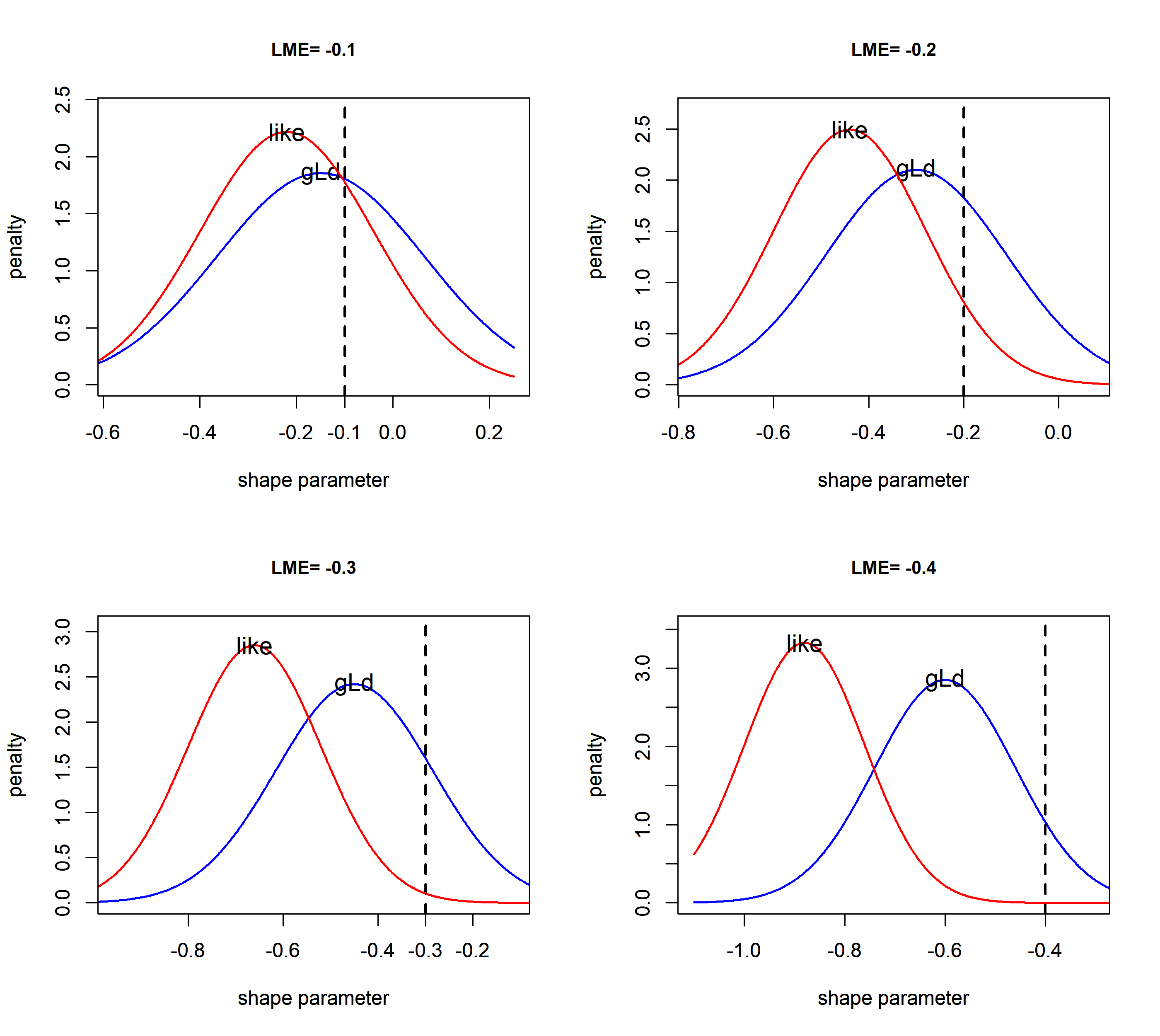} 
	\caption{Graphs of penalty functions which are different for weighting methods (like and gLd) and for $\hat \xi$ (LME). The vertical dashed line represents the LME obtained from data which are $-0.1,\, -0.2,\, -0.3$, and $-0.4$, respectively.}
	\label{fig:pen}
\end{figure}

{Figure~\ref{fig:pen} shows graphs of these PFs for weighting methods (MA.like and MA.gLd) and for different estimates of $\xi$ such as $-0.1,\, -0.2,\, -0.3$, and $-0.4$, respectively. As $\hat \xi$ decreases, a more pronounced shift to the left is assigned, so as to differentially correct the underestimation biases according to the value of $\hat \xi$.}

{Table \ref{sim_result.bma} shows the results of a simulation study using BMA with the above PFs, acquired under the same settings as in Section \ref{sec.simul}. Compared to Table \ref{sim_result}, considerable reduction of underestimation biases is achieved for $\xi \le -0.3$. However, new overestimation biases are observed for $\xi \ge -0.25$. The SE and RMSE of BMA.like are reduced from those of MA.like, and are comparable with or smaller than those of Re.MLE1 and LME. This result (the reductions of bias and SE) is a remarkable advantage of using BMA.like. Whereas, the SE and RMSE of BMA.gLd are increased relative to those of MA.gLd.}

\begin{table}[H]
	\centering
	\caption{Simulation results same as Table \ref{sim_result} but for Bayesian model averaging methods.
	}\label{sim_result.bma}
	\vspace{.3cm}
	\resizebox{\columnwidth}{!}{
		\begin{tabular}{c|c|cccccccccc}
			\hline
			Measure & Weight & -0.45 & -0.40 & -0.35 & -0.3 & -0.25 & -0.2 & -0.15 & -0.1 & -0.05 & 0.0  \\  \hline
			Bias & BMA.like &  -13.4 & -2.0 & 1.2 & 3.8 & 11.8 & 8.5 & 6.9 & 1.6 & 1.8 & -1.5 \\ 
			Bias & BMA.gLd &  -16.8 & -3.5 & -2.2 & 2.7 & 11.7 & 7.9 & 4.3 & -0.3 & -0.9 & -2.4 \\ \hline
			SE & BMA.like &  174.9 & 156.3 & 125.8 & 100.8 & 99.3 & 75.5 & 62.0 & 47.2 & 38.3 & 31.8  \\ 
			SE & BMA.gLd &  186.6 & 170.4 & 142.2 & 119.2 & 114.2 & 83.4 & 66.7 & 49.6 & 37.6 & 30.0  \\ \hline
			RMSE & BMA.like &  175.4& 156.3& 125.8& 100.9& 100.0&  76.0&  62.4&  47.2 & 38.3&  31.8 \\ 
			RMSE & BMA.gLd &  187.4& 170.4& 142.2& 119.2& 114.8&  83.8&  66.8 & 49.6&  37.6&  30.1 \\ 
			\hline
	\end{tabular}}
\end{table}

{Table \ref{tab:haenam_bma} provides the estimates and SEs calculated by BMA methods for Hae-nam data. Compared to Table \ref{Hae-nam_result}, the 100-year return levels are increased from MA methods. The `SE among model' and `SE within model' are calculated by taking the square roots of the first and second terms of the right hand side of the variance formula (\ref{bma-mean}), respectively. The `Total SE' is obtained by taking the square root of the variance formula (\ref{bma-mean}).
The (Total) SE of BMA.gLD increased from that of MA.gLd, whereas the SE of BMA.like is decreased from that of MA.like. In BMA.like, the among-model variance is relatively small compared to the within-model variance, which results in a decrease of total SE.  In contrast, the among-model variance in BMA.gLd is relatively large, which results in an increase of total SE.}

 {Although the BMA methods considered here showed a substantial improvement, we believe that more elaborate choices of PF and tuning constants could lead to more accurate estimation of return levels than the BMA method presented here.}


\begin{table}[ht]
	\caption{Same as Table \ref{Hae-nam_result} but for Bayesian model averaging methods. The `SE am(ong).model' and `SE wi(thin).model' are calculated by taking the square roots of the first and second terms of the right hand side of the variance formula (\ref{bma-mean}), respectively. The `Total SE' is obtained by taking the square root of the variance formula (\ref{bma-mean}).} \label{tab:haenam_bma}
	\vspace{.3cm} 
	\centering 
	\begin{tabular}{c|ccccccc}
		\hline
		Method& $\hat \mu$ & $\hat \sigma$ &$\hat \xi$ & 100-y RL & SE am.model & SE wi.model & Total SE \\  \hline
		BMA.like & 113.59 &  34.87& -0.348 & 520.59 & 14.73 & 60.29 & 62.07 \\
		BMA.gLd & 116.43 &  34.02&  -0.321 & 507.70 & 65.34 & 58.66 & 87.81 \\
	\hline
	\end{tabular}
\end{table}

{\subsection{Guide for using MA methods}
\begin{enumerate}
 \item [1.] {\bf Starter method: MLE}. The starter LME depends on the confidence interval obtained from the bootstrap, which may vary slightly across different runs. Whereas, the starter MLE rarely depends on this randomness.
	\item [2.] {\bf Selection of $K$}: Employ the method described in Section \ref{sec:selK}.
	\item [3.] {\bf Weighting criterion: MA.like1 or BMA.like}. In general, the MA.like1 seems to perform well in the simulation study. If the user is uncomfortable with trimming the lowest observation or prefers a Bayesian approach, the BMA.like method can be a good alternative.
\end{enumerate}
}

\section{Discussion}

\subsection{Bias correction}

{We considered the reasons why the MA methods proposed in this study underestimate the return level. One plausible answer can be found in the fact that the LME sometimes produces underestimation (UE) bias for some negative $\xi$ (especially for $\xi \le -0.3$). Consequently, the MA.like method results in UE, because it may accumulate UE by computing the LME $K$ times. In the  MA.gLd method, the covariance matrix $V$ was used in constructing $\text{GLD}_k$ in (\ref{GL-dist}) over all $k$, instead of using $V_k$. This uniform usage might cause a poor approximation to $\text{GLD}_k$, especially for large negative $\xi_k$. Further investigation is needed for this undesirable bias.} 
	
	{In addition, we employed the multivariate normal density (\ref{probMN}) to approximate the likelihood function of population L-moments given sample L-moments. This approximation may be valid for a large sample. But, it may not perform well for a small sample, which probably induces UE bias by assigning too small $p(\ux | M_k)$ values for large negative $\xi_k$. We may need to use an asymmetric distribution or the multivariate $t$-distribution instead of the normal density (\ref{probMN}).}


\subsection{Weighting scheme based on forward cross-validation} \label{sec:w_CV}

Contrary to the L-moments distance approach, many researchers have used cross-validation (CV) to determine weights for MA. Hansen and Racine (2012) proposed jackknife MA, where weights are selected based on leave-one-out CV. 
For time series MA, forward CV has been considered to preserve the temporal order in prediction (Hjorth 1982; Zhang \& Zhang 2023). 
Although the data in this study are assumed to be stationary, we applied the forward CV approach to  weight selection, as our primary goal is to predict high quantiles, mainly out-of-sample quantities (i.e., extrapolative predictions). 

To determine weights based on forward CV, we first rearrange the data in ascending order. We then hold out the largest $\alpha\%$ of values as a test set and use the remaining $(100-\alpha)\%$ of the data—referred to as the training set—to estimate the parameters of each submodel. For each submodel, we predict the test set value using the submodel trained on the training set and compute the forward CV (FCV) score as follows:
\beq \label{fcv}
FCV_k\ =\ \sum_{i=1}^{n_{te}}\ \frac{\{ y^{te}_i - \hat y^{te}_{i, k}  \}^2 }{ s^2_{i,k}},
\eeq
where the superscript $`te$' denotes the test set, $n_{te}$ is the number of observations in the test set, and $s^2_{i,k}= \text{Var}(\hat y^{te}_{i, k})$ represents the variance estimate of the prediction, obtained in Section \ref{sec:asvar}.
Here, $\hat y^{te}_{i, k}$ is computed using the plotting position estimation, where the parameters are estimated from the training set. 
 To compute $\hat y^{te}_{i, k}$, we consider the MLE and LME within each submodel $M_k$. A smaller value of $FCV_k$ indicates better extrapolative prediction performance, which is desirable when working with a limited number of extreme observations.
To assign a higher weight to submodels with smaller $\text{FCV}_k$ values, we adopted the following likelihood function of the normal distribution:
\beq \label{probCV}
p(\ux | M_k)\ =\   \prod_{i=1}^{n_{te}} \frac{1}{\sqrt{2 \pi }\, s_{i,k}} \times \text{exp} \left(- \frac{\text{FCV}_k}{2} \right) .
\eeq
Then, the weight for model $M_k$ is calculated using the same formula as in (\ref{weight}).
In addition to assigning weights using (\ref{probCV}), we also attempted to determine weights by minimizing (\ref{fcv}) through a numerical optimization routine. Zhang and Zhang (2023) established the asymptotic optimality of weights that minimize $\sum_{i=1}^{n_{te}}\ \left( y^{te}_i - \hat y^{te}_{i, k}  \right)^2 $ for time series data.

However, throughout our simulation study and real-data applications, we observed that the performance of the MA method with weights based on FCV is sensitive to the proportion of the training set (also referred to as the window size). We found that an appropriate proportion for reliable estimation depends on the return period and the tail heaviness. In this study, we could not systematically determine an optimal proportion that simultaneously accounts for both the return period and tail heaviness. Nevertheless, we believe that the FCV-based weighting is a promising criterion for weight assignment and should be further investigated in future research to enhance the accuracy of high-quantile predictions.

\subsection{Extension to other models}
Since it is known that the MA method generally performs well when the error variance is large (Liu et al.~2016, 2023), the proposed MA approach may be applicable to certain `sub-asymptotic' extreme value distributions that suffer from high error variance in high quantile estimation. One example is the four-parameter kappa distribution (Hosking 1994; Shin \& Park 2023; Strong et al.~2025), a generalization of the GEVD with an additional shape parameter. Given that the MLE for the four-parameter kappa distribution exhibits high estimation variance (Papukdee et al.~2022), the MA method may provide a more reliable estimator than the MLE.
  
Extending the proposed method to nonstationary (NS) extreme value models would also be beneficial. In NS extreme value modeling, the MLE is highly sensitive to outliers or a few influential observations toward the end of the sample, often leading to substantial estimation variance for high quantiles. To address this issue, some researchers (Strupczewski \& Kaczmarek 2001; Gado \& Nguyen 2016; Shin et al.~2025a) have proposed combining least squares and L-moment estimation methods. To fit an NS GEV model using the proposed MA approach, we first select $K$ submodels with fixed shape parameters. In NS GEV modeling, the shape parameter ($\xi_t$) is difficult to estimate precisely, so we treat $\xi_t$ as a constant. Under each fixed shape parameter $\xi_k$, we estimate the MLE of NS location and scale parameters. Next, the NS GEV submodel is transformed into a stationary extreme value model (Coles 2001). Under this transformed stationary model, the generalized L-moments distance is computed to assign weights to each submodel $M_k$. Further research is needed to explore this topic in greater detail.

\section{Conclusion}

We proposed a new MA method for estimating high quantiles of the GEVD. The proposed approach consists of selecting submodels based on the confidence interval of the shape parameter, estimating submodel parameters, and assigning weights to each submodel. Different criteria were employed for parameter estimation and weight assignment. Submodels were constructed using the MLE and LME, while weights were calculated using generalized L-moment distances and likelihood-based methods. This mixed approach, which leverages the strengths of both MLE and LME, is a key factor in improving accuracy in high quantile estimation. Additionally, we quantified the uncertainty of the MA estimator for return levels under a random-weight setting. Furthermore, a surrogate model for the MA method was developed for further analysis and applied to a real dataset--the annual maximum daily precipitation in Hae-nam, Korea.

Based on our simulation study, the proposed MA method with the `gLd' weighting scheme performed well for $-0.25 < \xi \le 0$, while the MA method with the `like' weighting scheme was effective for $\xi \le -0.25$. In general, without considering specific values of $\xi$ (i.e., tail heaviness), we recommend using the LME, the MA method with the `like1' weighting scheme, and the restricted MLE (or mixed estimator) considered by Morrison \& Smith (2002) and Ailliot et al.~(2011). 
{The Bayesian model averaging approach with a data-adaptive penalty function turned out to be a good alternative to existing estimation methods, especially for correcting the underestimation bias of the MA methods. We recommend using the BMA.like method as well.}

\renewcommand{\baselinestretch}{0.8}
\subsection*{Funding and Acknowledgments}
\begin{small}
	The authors are grateful to the reviewers and the associate editor of SERRA (Stoch Environ Res Risk Assess) for their valuable comments and constructive suggestions as these led to a greatly improved paper.
	We thank Prof.~Sanghoo Yoon for his help for this study.
	This work was supported by Basic Science Research Program through the
	National Research Foundation of Korea (NRF) funded by the Ministry of Education (RS-2025-25436608).
	
	\subsection*{Code and data availability}
	Maximum rainfall data at Hae-nam, Korea: https://github.com/yire-shin/MA-gev/tree/main/data  \\
	R code for model averaging: https://github.com/yire-shin/MA-gev.git
	
	\subsection*{Conflict of interest}
	The authors declare no potential conflicts of interest.
	
	\subsection*{ORCID}
	
	Yire Shin, 0000-0003-1297-5430; $~~$ Yonggwan Shin, 0000-0001-6966-6511 \\
	Jeong-Soo Park, 0000-0002-8460-4869 
	
	\subsection*{Author contributions statement}
	 Yire and Park conceived the study, Yonggwan and Yire conducted the analysis. All authors wrote the first draft, reviewed, edited, and approved the final manuscript.
	
\end{small}

\begin{small}
\renewcommand{\baselinestretch}{0.8}

	\end{small}

\end{document}